\documentclass[12pt]{article}
\usepackage{amsthm,amsmath,amsfonts,amssymb}
\usepackage{graphicx,epsfig}
\numberwithin{equation}{section}
\newcommand{\R}{{\mathord{\mathbb R}}}

\newcommand{\N}{{\mathord{\mathbb N}}}
\newcommand{\C}{{\mathord{\mathbb C}}}

\newcommand{\HH}{\mathcal{H}}

\newcommand{\JJ}{\mathcal{J}}
\def\tr{\operatorname{tr}}

\def\supp{\operatorname{supp}}
\def\e{{\rm e}}
\def\i{{\rm i}}
\def\dn{|\hskip -1.2pt|\hskip -1.2pt|}
\def\Im{\operatorname{Im}}

\def\ad#1#2#3{{\operatorname{ad}}^{(#1)}_{#2}({#3})}
\def\O#1{{\mathrm{O}(#1)}}
\def\o1{{\mathrm{o}(1)}}
\def\slim{\mathop{\mathrm{s-}}\!\lim}
\def\Lim{\mathop{\mathrm{Lim}}}
\newtheorem{thm}{Theorem}[section]
\newtheorem{lem}[thm]{Lemma}

\newtheorem{prop}[thm]{Proposition}

\title{Adiabatic charge pumping in open quantum systems}
\author{\hspace{-.2 cm} J.E. Avron${}^{(a)}$, A. Elgart${}^{(b)}$, G.M. Graf${}^{(c)}$,
L. Sadun${}^{(d)}$, K. Schnee${}^{(e)}$\\
\normalsize\it \hspace{-.5 cm}\\
\hspace{-.5 cm}\normalsize\it ${}^{(a)}$ Department of Physics,
Technion, 32000 Haifa, Israel\\ ${}^{(b)}$\normalsize\it 
Courant Institute,  New York University,  New York, NY 10012, USA 
\\
\normalsize\it \hspace{-.5 cm}${}^{(c)}$ Theoretische Physik,
ETH-H\"onggerberg,
8093 Z\"urich, Switzerland\\
\normalsize\it \hspace{-.5 cm}${}^{(d)}$ Department of
Mathematics,
University of Texas, Austin, TX 78712, USA\\
\normalsize\it \hspace{-.5 cm}${}^{(e)}$ Department of
Mathematics, Caltech, Pasadena, CA 91125, USA}

\begin{document}
\maketitle
\vspace{0.4cm}
\begin{abstract}
We introduce a mathematical setup for charge transport in
quantum pumps connected to a number of external leads. It is proved that
under rather general assumption on the Hamiltonian describing the
system, in the
adiabatic limit, the current through the pump is given by a formula of
B\"uttiker, Pr\^etre, and Thomas, relating it to the frozen
$S$-matrix and its time derivative.
\end{abstract}
\input epsf
\section{Introduction}\label{INTRO}
Transport in quantum pumps has been investigated in relation to various
properties and from many perspectives \cite{BPT, B, AA, SAA, IL2, L3}.
The goal of this article is to
provide a rigorous setting for a single but important aspect of these
devices, namely the charge transport or, more precisely, its expectation
value. The idealized setting is as follows: a pump, whose internal
configuration varies slowly in time in a prescribed manner, is connected
to $n$ leads, or channels, along each of
which independent electrons can enter or leave the pump.
We assume that the electron in the lead has no transverse or spin
degrees of freedom and may be thought of as a (non-relativistic)
particle moving on a half line\footnote{Such extra degrees of
freedom can be represented by adding channels. In general, the
different channels may then have different propagation speeds. }.
The incoming electron distribution, at zero temperature,
 is a Fermi sea with
Fermi energy $\mu$ common to all leads. As a rule, this does not apply
to the distribution of the outgoing electrons, as their energies may
have been shifted while scattering at the pump. Because of this imbalance
a net current is flowing in the leads. The expected charge
transport is expressed by the formula \cite{BPT, B}
\begin{equation}
dQ_j= \frac{e}{2\pi} (\i (dS)S^*)_{jj}\;.
\label{eq:bpt}
\end{equation}
Here $S=(S_{ij})$ is the $n\times n$ scattering matrix at energy $\mu$
computed as if the pump were {\it frozen} into its instantaneous
configuration. A change of the configuration is accompanied by a
change $S\to S+dS$ of the scattering matrix and by a net charge
$dQ_j$ leaving the pump through lead $j$. Finally $e$ is the
electron charge, which is henceforth set equal to $1$. \\

We shall next present a mathematical framework in which (\ref{eq:bpt})
can be phrased as a theorem.
\begin{list}{$\bullet$}{\setlength{\leftmargin}{\parindent}}
\item
The {\it single-particle Hilbert space} is
given as
\begin{equation}
\HH = \HH_0 \oplus L^2(\R_+,\C^n)\;,
\label{eq:hs}
\end{equation}
where states in $L^2(\R_+,\C^n)=\oplus_{j=1}^n L^2(\R_+)$, resp.~in
$\HH_0$, describe an electron in one of the leads
$j=1,.\ldots n$, resp.~in the
pump proper. The latter Hilbert space is not further specified, but the
hypothesis A2 below confers on the pump the role of an abstract finite
box \cite{SZ}. Let
$\Pi_j:\HH \to\HH$ denote the projection onto
$\HH_0$ for $j=0$ and on $j$-th copy of $L^2(\R_+)$ for
$j=1.\ldots n$.
\item
Since the pump configuration is supposed to change slowly in time $t$,
we will eventually consider the evolution of the electrons in an
adiabatic limit, where $s=\varepsilon t$ is kept fixed as
$\varepsilon>0$ tends to $0$. In terms of the rescaled time coordinate
$s$, called {\it epoch}, the propagator $U_\varepsilon(s,s')$ on
$\HH$ satisfies the non-autonomous Schr\"odinger equation
\begin{equation}\label{SCHRO_EQ}
\i \partial_s U_\varepsilon(s,s') =
\varepsilon^{-1} H(s) U_\varepsilon(s,s')\;,
\end{equation}
where $H(s)$ is a family of self-adjoint {\it Hamiltonians} on
$\HH$ enjoying the following properties:
\begin{eqnarray*}
\hbox{(A1)}&&\qquad H(s) -H(s')\hbox{ is bounded and smooth in $s$}\;, \\
\hbox{(A2)}&&\qquad \|(H(s)+\i)^{-m} \Pi_0\|_1< C \hbox{ for all $s$ and some
$m\in \mathbb{N}$}\\
\hbox{(A3)}&&\qquad H(s) \psi = -d^2\psi/dx^2 \hbox{ for }
\psi \in C_0^\infty(\R_+,\C^n)\;,\\
\hbox{(A4)}&&\qquad\sigma_{\rm{pp}}(H(s)) \cap (0,\infty) = \emptyset \;,\\
\hbox{(A5)}&&\qquad H(s) = H_- \hbox{ for } s \le 0 \;.
\end{eqnarray*}
Here $\|\cdot\|_1$ denotes the trace class norm over $\HH$, while the
operator norm will be written as $\|\cdot\|$.
Assumption A3 states that a particle in the leads is free; in
particular, together with A1, it implies that changes in the
Hamiltonian are confined to the pump proper:
\begin{equation}
\qquad H(s)-H(s') = (H(s)-H(s'))\Pi_0\;.
\label{eq:conf}
\end{equation}
Assumption A4 requires that there are no positive embedded
eigenvalues; A5 states that the pump is at rest for $s \le 0$.

\item The initial state of the electrons should be
an equilibrium state. This is achieved thanks to Assumption A5 by
positing that the {\it 1-particle density matrix} at some (and hence any)
epoch $s_-<0$ is of the form $\rho(H_-)$, where $\rho(\lambda)$ is a
function of bounded variation with $\supp d\rho\subset(0,\infty)$. A
good example is the Fermi sea, where $\rho(\lambda)=\theta(E-\lambda)$.
The time evolution then acts as
\begin{equation}
\rho(H_-)\mapsto U_\varepsilon(s,s_-)\rho(H_-)U_\varepsilon(s_-,s)\;.
\label{eq:dm}
\end{equation}

\item We define a generator of exterior scaling w.r.t.
(\ref{eq:hs}) by
\begin{equation}
A=0\oplus\frac{1}{2\i}\left(\frac{d}{dx}v(x)+v(x)\frac{d}{dx}\right)\;,
\label{eq:gdil}
\end{equation}
where $v(x):[0,\infty)\to \R$ is smooth with $v(x)=0$, resp.
$=x$ for small, resp. large $x$, and $v'(x)\ge 0$ everywhere. We
note that $A=A^*$ commutes with $\Pi_j$, and set $A_j=A\Pi_j$. The
operator $A$ distinguishes between {\it incoming} and {\it outgoing} states,
respectively associated with spectral subspaces $A<-a$ and $A>a$
with some large $a>0$. Detection of a particle, and hence of its
charge, deep inside lead $j$ may be realized as the operator
$Q_j(a)=f(A_j-a)+f(-A_j-a)$, where $f\in C^\infty(\R)$ is a switch
function: $f(\alpha)=0$ for $\alpha<-1$, resp. $=1$ for
$\alpha>1$. The {\it current operator} then consistently is
\begin{equation}
I_j(a)=\i[H(s), f(A_j-a)+f(-A_j-a)]=:I_{j+}(a)+I_{j-}(a)\;.
\label{eq:cu}
\end{equation}
(In what follows we shall sometimes suppress the index $j$ for the
sake of notational simplicity). One feature of this choice
of current operator is that the ``ammeter'' is located not at a
fixed distance from the pump, but rather at a fixed number of
wavelengths, $a$, from it: The longer the wavelength the more
distant the ``ammeter''. In the case that one focuses on a narrow
energy interval, say near the Fermi energy, the ``ammeter'' is
also at essentially fixed distance from the pump.

We remark that by the support
property of $v$ and by A3, the above commutator does not depend
on $s$. The expectation value of the current at epoch $s$, i.e.,
in the state (\ref{eq:dm}), is then given as
\begin{equation}\label{current}
\langle I\rangle_j(s,a,\varepsilon)=
\tr\big(U_\varepsilon(s,s_-)\rho(H_-)U_\varepsilon(s_-,s)I_j(a)\big)\;.
\end{equation}
\end{list}
In contrast to $Q_j(a)$, which clearly has an infinite expectation value
in that state, $I_j(a)$ is inclined to have a finite one. Moreover, it
should behave as $\varepsilon$ if, in accordance
with (\ref{eq:bpt}), the charge $dQ_j$ transferred during
$ds=\varepsilon^{-1}dt$ is to have a non trivial limit as
$\varepsilon\to 0$.\\

Other realizations of the current operator are possible\footnote{The canonical
choice of the current operators one normally finds in textbooks corresponds to
$f(x)=\theta(x)$.}, such as
$\i[H(s), f(x_j-a)]$, or the example based on the precession of a spin
proposed in \cite{L3}, and the result
(\ref{eq:bpt}) should be independent of the choice. Our definition
(\ref{eq:cu}) has the property that $I_j(a)$ naturally splits into two parts
distinguished by their Heisenberg dynamics: The outgoing current $I_{j+}(a)$
which is essentially free in the future and the incoming current $I_{j-}(a)$
which is free in the past. As the initial condition is set in the past of the
measurement, only $I_{j+}(a)$ will be affected by scattering.
\begin{figure}
\hskip 1.3 in \includegraphics[width=10cm]{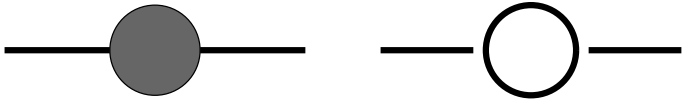}
\caption{A
scatterer with two channels (left) and its disconnected analog
(right). Neumann boundary conditions are imposed at the edges of
the channels. } \label{scattererfig}
\end{figure}
\begin{list}{$\bullet$}{\setlength{\leftmargin}{\parindent}}
\item
Finally we ought to state the reference dynamics $H_0$ to which $H(s)$ is
compared in the definition of the (frozen) {\it scattering operator}
\cite{RS3, Y}
\begin{equation}
S(s)=S(H(s),H_0)\;,
\label{eq:scatt}
\end{equation}
whose fibers $S(s,E)$, called {\it scattering matrices},
appear in (\ref{eq:bpt}). While it turns out that
the choice of $H_0$ is largely irrelevant, for the sake of simplicity let
$H_0$ be the Laplacian, acting on $L^2(\R_+,\C^n)$, with a Neumann boundary
condition at $x=0$. $S(s,k^2),\,(k>0)$, then agrees with the familiar
definition based on generalized eigenfunctions incident through lead $i$:
$(\delta_{ji}\e^{-\i kx}+S_{ji}\e^{\i kx})_{j=1}^n$. In particular, this
reproduces the standard form of the scattering matrix of two channels:
$$S=\begin{pmatrix}
  r & t' \\
  t & r'
\end{pmatrix}
$$ with $r,\,r'$ the right and left reflection amplitudes and $t,\,t'$ the
corresponding transmission amplitudes.
\end{list}
The fundamental equation (\ref{eq:bpt}) may thus be given the following
reformulation:
\begin{equation}
\lim_{a\to\infty}\lim_{\varepsilon\to 0} \varepsilon^{-1}\langle
I\rangle
_j(s,a,\varepsilon) =-\frac{\i}{2\pi}\int_0^\infty
d\rho(E)\Bigl(\frac{d S}{d s}S^*\Bigr)_{jj}\;. \label{eq:bpt1}
\end{equation}
In particular, for the Fermi sea $\rho(H_-)=\theta(\mu-H_-)$ as initial
state, we recover (\ref{eq:bpt}) from $d\rho(E)=-\delta(E-\mu)dE$. The
limit $a\to\infty$ is taken so as
to have the current measurement made well outside of the scattering
region, but after the adiabatic limit
$\varepsilon\to 0$. By doing so, a current is still measured within the same
epoch as the scattering process which generated it, though at
a different time.\\

The result can be proven essentially in this form, though some problems 
that arise both in the infrared and the ultraviolet have to be dealt 
with. In fact, the adiabatic limit is realized
\cite{AEGS3} in the regime where the dimensionless quantity given
by $\varepsilon$ times the dwelling time of an electron in the
pump is small. Low energy particles have a large dwell time in the
pumps and, in addition, may get trapped indefinitely as new bound
states are born at the threshold $E=0$. This means that no
scattering description in terms of a single epoch is adequate at
low energies. Similarly, at high energies resonances may become
increasingly sharp with correspondingly long dwelling times
\footnote{If the pump is a chaotic billiard, the classical
dynamics will, in general, have arbitrarily long periodic orbits.
It is natural to expect that such orbits give rise to
resonances.}. On the other hand, low, resp. high energy states do
not contribute to the net current, since they are filled, resp.
empty in both the incoming and the outgoing flow. We shall
therefore concentrate on the contribution to the current coming
from states in any intermediate energy range, as selected by the
function $\chi$ below.
\begin{thm}\label{BPT}
Let $\chi\in C_0^\infty(0,\infty)$ with $\chi=1$ on $\supp d\rho$.
Redefine the current operator (with a UV and IR cutoff)
\begin{equation}
I_{j,\pm}(s,a)=
\chi\big(H(s)\big)\i\,[H(s), f(\pm A_j-a)]\chi\big(H(s)\big)\;,\qquad(a>1)\;,
\label{eq:cu1}
\end{equation}
in (\ref{eq:cu}). Then
\begin{equation}
\lim_{a\to\infty}\lim_{\varepsilon\to 0} \varepsilon^{-1}\langle
I\rangle_j(s,a,\varepsilon) =-\frac{\i}{2\pi}\int_0^\infty
d\rho(E)\Bigl(\frac{d S}{d s}S^*\Bigr)_{jj}\;, \label{eq:bpt2}
\end{equation}
where $S=S(s,E)$. The double limit is uniform in $s\in I,\, I$
being a compact interval, whence it carries over to the
transfered charge $\int_0^{s/\varepsilon}dt'\langle
I\rangle_j(\varepsilon t',a,\varepsilon)=
\varepsilon^{-1}\int_0^sds'\langle I\rangle_j(s',a,\varepsilon)$.
\end{thm}
\noindent{\bf Remark.}
The auxiliary objects $v$ and $f$ affect
the current operator. Nevertheless, they disappear from its expectation value
in the adiabatic and the large $a$ limits. This may be phrased as the
statement that different ammeters measure the same current. \\

We first give a heuristic derivation of Eq.~(\ref{eq:bpt2}), which may 
however serve as a guide through the complete proof. (For more hints, 
see \cite{AEGS3}.) One could argue that for computing the 
expectation values of observables, which, like $I_{j,-}(s,a)$, pertain 
to the incoming part of phase space, scattering may be ignored, i.e., one 
may pretend that the state of the system at epoch $s$ simply is 
$\rho(H(s))$. On the other hand, when discussing $I_{j,+}(s,a)$ one should do
as if the state were 
\begin{equation}\label{eq:rhoout}
\rho(H(s))+\varepsilon[S^{(1)}(s),\rho(H(s))]+\ldots\;,
\end{equation}
where 
\begin{equation}\label{eq:dscatt0}
S^{(1)}(s)=-\i \int_{-\infty}^{\infty} dt\, t\, \e^{\i H(s)t}
\dot{H}(s)\e^{-\i H(s)t} \;.
\end{equation}
This comes from linearizing the Hamiltonian
$H(s+\varepsilon t)=H(s)+\varepsilon \dot H(s) t+\ldots$ around epoch $s$,
since $U=1+\varepsilon S^{(1)}(s)+\ldots$ is then the propagator from
$t=-\infty$ to $t=\infty$ in the interaction picture based on $H(s)$. Now 
(\ref{eq:rhoout}) is the state $U\rho(H(s))U^*$ after scattering. 
The contribution of the first term there to 
$\langle I\rangle_{j,+}$ cancels against $\langle I\rangle_{j,-}$, see 
Lemma~\ref{CURR_STAT}. The contribution of second term is responsible for the
r.h.s. of Eq.~(\ref{eq:bpt2}). To foreshadow this outcome we formally rewrite
it, see Lemma~\ref{sc5}, as
\begin{equation}
[S^{(1)}(s),\rho(H(s))] =
-\i\partial_{s'}S(s',s)_{s'=s}\rho'(H(s))\;,
\label{eq:s1}
\end{equation}
with
\begin{equation}
\partial_{s'}S(s',s)_{s'=s}=
-\i \int_{-\infty}^\infty dt\, \e^{\i H(s)t}\dot{H}(s) \e^{-\i H(s)t}\;.
\label{eq:s2}
\end{equation}
Here, $1+(s'-s)\partial_{s'}S(s',s)_{s'=s}+\ldots$ is the Born approximation
for the scattering matrix $S(s',s)$ for the pair of autonomous Hamiltonians 
$(H(s'),H(s))$. The importance of Eq.~(\ref{eq:s1}) is twofold. First 
it reduces matters to ``frozen'' scattering data and their derivatives. 
Second, it makes it clear why only the variation $d\rho(E)$ matters and, in
particular, why for the Fermi sea
$\rho(\lambda)=\theta(\mu-\lambda)$ only states at the Fermi energy $\mu$
contribute to the current.\\

Let us next discuss some of the conceptual issues involved in this result.

1) Theorem~\ref{BPT} 
plays the role of the Kubo formula and like it gives a handle on transport,
including dissipative transport, using Hamiltonian evolution. A well known
weakness of linear response theory, stressed by van Kampen \cite{vK}, is 
that it takes the limit of linear response first and only then the 
thermodynamic limit. The correct order is, of course, the reverse. 
Theorem~\ref{BPT} is free of this criticism
in that one starts with a system that has infinite extent, for which it is
nevertheless possible to estimate the error made by linearizing the dynamics,
see Lemma~\ref{MAIN_S4}. The reason for that is the
finite extent of the pump or, more precisely, the finite dwell times. 

2) Theorem~\ref{BPT} is an example of an adiabatic theorem for open, gapless
systems. In contrast to other results of that kind, such as \cite{Bo, AE}
which establish that an embedded eigenstate evolves adiabatically, ours
is about the evolution associated with an infinite dimensional
spectral subspace, e.g. the Fermi sea. Moreover, the goal is not
to establish that a spectral subspace is preserved by the dynamics; rather,
it is to determine the amount by which it mingles with its complement, and
the current is a measure thereof.

3) On the l.h.s. of (\ref{eq:bpt2}) the scattering process and the 
measurement of the current are described as a single quantum history. This
should be contrasted to usual textbook treatments of scattering rates, which
have the following features: (i) The rates are computed classically as the 
product of a quantum mechanical scattering cross section and of an 
incident current; (ii) the use of the scattering cross section tacitly 
replaces the actual state by its free asymptote. For a static potential
and a steady beam both steps have been justified \cite{DF} in the limit 
where the measurement is deferred all the way to $t\to\infty$. This is 
not quite so in the present fluctuating setting, where the ammeter is 
located at an intermediate scale, as specified by the order of the 
$\varepsilon,\, a$ limits: far away from the pump on 
the scale of the scatterer, but close to it on the scale of the distance
traveled by a particle on the adiabatic time scale. 

4) The formula of B\"uttiker et al., Eq.~(\ref{eq:bpt}), is pointwise in
time. Suppose we represent the flow by means of coherent states of widths 
$\Delta E$ and $\Delta t$. In line with the remark just made, one might
suspect that $\Delta t$ needs to be picked small w.r.t. to the adiabatic time
scale $\varepsilon^{-1}$, and hence $\Delta E\gtrsim \hbar\varepsilon$ by
the Heisenberg uncertainty relation. At
temperature $\beta^{-1}$ the 1-particle distribution of the incoming flow is 
$(1+e^{\beta(E-\mu)})^{-1}$, which suggests that 
$\Delta E\lesssim \beta^{-1}$. This seemingly presents an obstacle to applying
scattering arguments when $\beta^{-1} < \hbar \varepsilon$.  Our result, which
applies to $\rho(\lambda)=\theta(\mu-\lambda)$, 
shows that this obstacle can be overcome, and that one {\it can} take
the zero temperature limit  $\beta\to \infty$
before taking $\varepsilon\to 0$. \\

Results related to Theorem~\ref{BPT} have been obtained at the level of both 
physical and mathematical rigor.
Among the former we mention, besides of \cite{BPT, B}, the 
formula of Lee, Lesovik and Levitov \cite{L3}, which expresses the 
noise generated by quantum pumps (as well as other moments) in terms of 
the scattering data. However, the relations regarding noise are
not local in time. Rather these are integral relations
that hold for a cycle of the pump. Physically, what sets the current 
apart from the noise is its linear dependence on $\rho$.
Rigorous mathematical description of the adiabatic scattering 
of wave packets is discussed in \cite{MN, MS}. Like our
result, they rely on propagation estimates. \\

The plan of the rest of the paper is as follows. In
Sect.~\ref{sect:2} we shall verify that
$\langle I\rangle_j(s, a,\varepsilon)$ and the Stieltjes integral
in (\ref{eq:bpt2}) are well-defined.
Further preliminaries, like propagation estimates, will be addressed in
Sect.~\ref{sect:2a}. In Sect.~\ref{sect:3} we shall compute the limit
$\varepsilon\to 0$ of the current in terms of data involving $H(s)$ and 
$\dot H(s)$, but oblivious of the past $\{H(t)\}_{t<s}$. The expression 
will further
reduce to the ``frozen'' scattering data (\ref{eq:scatt}) in
Sect.~\ref{sect:4} where the limit $a\to\infty$ is
taken. In the Appendix we establish some trace class estimates for
operators related to (\ref{eq:cu1}). The main
ideas are further discussed at the beginning of 
Sects.~\ref{sect:3}, \ref{sect:4}.\\

We conclude with a remark on notation. Multiple commutators
are denoted by $\ad{k}{A}{B}= [\ad{k-1}{A}{B},A]$ with
$\ad{0}{A}{B}=B$. By $F(A\ge a)$ we mean the spectral projection of $A$
onto $[a,\infty)$. The trace class ideal is denoted as $\JJ_1$.
Generic constants are indicated by $C$.

\section{The current operator and the scattering matrix}\label{sect:2}
The state of independent quantum particles is described by a
density matrix $0\le\rho$ normalized so that $\tr \rho$ is the
number of particles. In the case of Fermions, $\rho\le 1$.
Thermodynamic systems have $\tr\,\rho=\infty$ and observables
that are otherwise innocent, and bounded operators in particular,
may fail to have finite expectation values. For example, the
charge associated with the ``box'' $f(A-a)$ in phase space is
infinite. Nevertheless, the current flowing into the box should
have a finite expectation. We begin by showing that the incoming
and outgoing current operators of Eq. ~(\ref{eq:cu}) are trace
class and consequently, the expectation value
$\langle I\rangle_j(s,a,\varepsilon)$ in any Fermionic state is well defined.

A classical interpretation of the trace class condition is for
the observable to be associated with a localized (bounded)
function in phase space. The heuristic reason why the current is trace
class is then as follows: The commutator $[H(s), f(A_j-a)]$ is
localized near the boundary of the box: a curve (hyperbola) in
phase space. This is where the ammeter is. The ultraviolet and
infrared cutoff $\chi(H(s))$ then further delineate a compact
region of phase space near the hyperbola. Our first preliminary result 
confirms this picture.
\begin{prop}\label{Itraceclass}
The operator of incoming and outgoing current in the $j$-th channel,
$I_{j\pm}(s,a)$, is a trace class operator which is localized near
$\pm a$ in the sense that
\begin{equation}
\Vert F(|A\mp a|\ge\alpha)I_{j\pm}(s,a)\Vert_1 \le C_N(1+\alpha)^{-N}
\end{equation}
for all $N\in\mathbb{N},\,\alpha\ge 0$.
\end{prop}

\begin{proof}[{\bf Proof.}]
This proposition is a direct consequence of Lemma~\ref{SECONDL_PRE} below.
\end{proof}
The pumping formula of Thm.~\ref{BPT} implies that no current flows 
if the pump is not operating, since $dS/ds=0$ if $H(s)$ is independent 
of $s$. Indeed, in the state $\rho(H(s))$, like in a
thermal state, 
different leads are at equilibrium with one another; moreover, in each lead,
right and left moving states yield compensating currents if equally occupied. 
When the pump is in operation, it will still prove useful to know that no 
persistent currents are flowing in the ``adiabatic'' state $\rho(H(s))$.
\begin{lem}\label{CURR_STAT} The currents in the state $\rho(H(s))$ vanish, 
namely
\begin{equation}
 \tr \big( \rho(H(s))(I_{j+}(s,a)+I_{j-}(s,a))\big) = 0\;.
\label{eq:stat}
\end{equation}
\end{lem}

\begin{proof}[{\bf Proof.}]
We omit $s$ from the notation. Since $I_\pm(a)\in\JJ_1$ it suffices to
prove (\ref{eq:stat}) for smooth $\rho$, since one may approximate the
general case by a sequence $\slim_{n\to\infty}\rho_n(H)=\rho(H)$.\\

First we show that $\tr (\rho(H)I_\pm(a))$ is independent of $a$.
Note that
$I_\pm(a_1) - I_\pm(a_2) =\chi(H) \i[H,g(A_j)]\chi(H)$ with
$g(A_j) =f(\pm A_j-a_1)-f(\pm A_j-a_2)$. Since
$\chi(H)g(A_j)=\chi(H)g(A)\Pi_j \in \JJ_1$,
see (\ref{eq:a6}), we have
\begin{equation}
\tr(\rho(H)(I_\pm(a_1)-I_\pm(a_2))) =
\i \tr (\rho\chi(H)H g(A_j)\chi(H) -
\rho\chi(H)g(A_j)H\chi(H)) = 0
\label{eq:indep}
\end{equation}
by cyclicity. It thus suffices to prove (\ref{eq:stat}) in the limit of
$a \to \infty$. To this end, let $H_0$
be the Neumann Hamiltonian on the leads introduced
below Eq.~(\ref{eq:scatt}), and let $J: L^2(\R_+,\C^n)\to \HH$ be
the embedding given by (\ref{eq:hs}). We maintain that
\begin{equation}
\lim_{a \to \infty}\|(\rho(H)I_\pm(a) -J\rho(H_0)I^0_\pm(a)J^* \|_1 = 0\;,
\label{eq:diffc}
\end{equation}
where $I^0_\pm(a)$ is defined as in (\ref{eq:cu1}) with $H$ replaced by
$H_0$. Indeed, by expanding that commutator we reduce matters to two
estimates, both of the form
\begin{equation*}
\lim_{a \to \infty}\|\chi_1(H)f(\pm A_j-a)\chi_2(H)
-J\chi_1(H_0)f(\pm A_j-a)\chi_2(H_0)J^*\|_1 = 0\;,
\end{equation*}
with $\chi_i\in C_0^\infty(\R),\,(i=1,2)$. Then we write the
difference as
\begin{equation*}
(\chi_1(H)-J\chi_1(H_0)J^*)f(\pm A_j-a)\chi_2(H)
+J\chi_1(H_0)J^*f(\pm A_j-a)(\chi_2(H)-J\chi_2(H_0)J^*)\;.
\end{equation*}
Since $\chi_i(H) -J\chi_i(H_0)J^*\in J_1$ by (\ref{eq:a2}) and
$\slim_{a\to\infty}f(\pm A_j-a)=0$, Eq. (\ref{eq:diffc}) follows. At
this point we only need
\begin{equation*}
\tr(\rho(H_0)(I^0_+(a)+I^0_-(a)) = 0\;,
\end{equation*}
which holds by time reversal invariance: For $K\psi(x)=\bar\psi(x)$ we have
$KH_0K=H_0,\,KAK$ $=-A$ and hence $KI^0_\pm(a)K=-I^0_\mp(a)$.
\end{proof}
\noindent{\bf Remark.}
One can establish the result without explicitly using time
reversal and using instead the pull through formula
Eq.~(\ref{eq:a4}).\\

Mourre theory, see e.g.~\cite{ABG}, plays a double role in our
analysis. First, it is at the heart of time dependent scattering
theory and the propagation estimates that we shall discuss in the
next section. At the same time, the theory also plays a role in
time independent methods and we shall use it to establish the
differentiability of the scattering matrix which appears
on the right hand side of Eq.~(\ref{eq:bpt2}).

\begin{prop}
Under the Assumptions A1-A5, the fibers of the frozen S matrix,
$S(s,E)$, are continuously differentiable in $s$ at $E>0$. In particular,
the integral on the right hand side of Eq.~(\ref{eq:bpt2}) is well
defined for $\rho$ with bounded variation.
\end{prop}
\begin{proof}[{\bf Proof.}]
The Hamiltonians $H(s)$ are dilation
analytic of type (A) with respect to the conjugate operator $A$. In
particular,
\begin{equation}
\ad{k}{A}{H}\hbox{ is } H(s)\hbox{-bounded.}
\label{eq:mc}
\end{equation}
More importantly, for any
energy $E>0$ the Mourre estimate holds
\begin{equation}\label{MOURRE}
E_{\Delta}(H(s)) \i[H(s),A]E_{\Delta}(H(s)) \ge \theta_0 E_{\Delta}(H(s))
\end{equation}
for some open interval $\Delta \ni E$ and $\theta_0 >0$. Note that a
compact term on the r.h.s can be dismissed by Assumption A4. The
Mourre estimate (\ref{MOURRE}) is stable under small bounded perturbations of
$s$, which is seen from (\ref{eq:a1}) and from
the fact that $\i[H(s),A]$ is independent of $s$, just as the commutator
in (\ref{eq:cu}). Therefore, (\ref{MOURRE}) holds uniformly in $s \in I$
and $E \in J$, with $J \subset (0,\infty)$ compact, and so do the
usual consequences of these assumptions. They comprise:
\begin{itemize}
\item[i)] Resolvent smoothness. Let
$\langle A\rangle = (1 + A^2)^{\frac{1}{2}}$ and $r > 1/2$. Then
\begin{equation*}
B(z,s) =
\langle A\rangle^{-r}(H(s)-z)^{-1}\langle A\rangle^{-r}
\end{equation*}
has smooth boundary values at $z=E+\i0, (E \in J)$, satisfying
\begin{equation}\label{JMP_EST}
\| \partial^k_E B(E+\i0,s)\| \le C_k
\end{equation}
for $k=0,1,\ldots$, see \cite{JMP}. Since
\begin{equation*}
\partial_s B(z,s) =
-\langle A\rangle^{-r}(H(s)-z)^{-1}\dot{H}(s)(H(s) -z)^{-1}
\langle A\rangle^{-r}
\end{equation*}
and $\dot{H}(s) = \langle A\rangle^{-r}\dot{H}(s) \langle
A\rangle^{-r}$,
the function $B(E+\i0,s)$ is jointly continuously differentiable in
$(E,s)$.
\item[ii)] $H$-smoothness. $\langle A\rangle^{-r}$ is
$H(s)$-smooth \cite{RS3, Y}, as a consequence of (\ref{JMP_EST}) for $k=0$.
\item[iii)] Stationary representation of the scattering matrix. Let
\begin{equation}\label{HILB_DECOMP}
E_J(H(s))\HH \to \int_J^\oplus \C^n dE\;, \qquad
\psi \mapsto \{\psi(E)\}
\end{equation}
be the spectral representation for $H(s)$ on $J$, and set
$\Gamma_0(E)\psi = \psi(E)$. Then 
$\Gamma_0(E)\langle
A\rangle^{-r}: \HH \rightarrow \C^n$ is bounded by (ii). Let
$S(s',s)$ be the scattering operator for the pair $(H(s'), H(s))$. Its
fibers $S(E): \C^n \rightarrow \C^n$ with respect to (\ref{HILB_DECOMP})
admit the representation \cite{Y}
\begin{equation}
S(E) =
1 - 2 \pi \i
\Gamma_0(\lambda)(V-V(H(s')-(E+\i 0))^{-1}V)\Gamma_0(\lambda)^* \;,
\label{eq:strep}
\end{equation}
with $V=H(s')-H(s)$. The r.h.s. is defined pointwise because of
$V=\langle A\rangle^{-r} V \langle A\rangle^{-r}$ and of (i). From this,
Eq.~(\ref{eq:strep}), and the statements about $B(E+\i0,s)$ we see
that $S(s',s)E_J(H(s))$ is continuously differentiable in $s' \in I$,
and that the derivative can be computed fiber-wise. The same applies to the
scattering operator $S(s)=S(H(s),H_0)$, though (\ref{eq:strep}) appears
slightly modified.
\end{itemize}
\end{proof}

\section{Propagation estimates}\label{sect:2a}
Propagation estimates play a key and multiple role in our
analysis. One role is that they guarantee that particles do not get
stuck in the pump. Consequently, the scattered particle indeed sees a
frozen scatterer to lowest order in the adiabatic limit, and a
linearly changing scatterer to first order.  At the same time
propagation estimates also play a role in establishing that the
current measured by the ammeter at epoch $s$ is determined by the
state of the pump at the same epoch, in the adiabatic limit.  This may
be interpreted as a statement that the particles neither linger nor
disperse too badly in the channels \footnote{Dispersion arises because
we use a non-relativistic (quadratic) kinetic energy in the leads. In
much of the literature on pumps dispersion is circumvented by making
the kinetic energy linear in the momentum.}. Propagation estimates
play yet another role in establishing the ``rigidity'' of the current:
The expectation value for the current a-priori depends on a choice of
a switch function $f$, the function $v(x)$ in the generator of
dilation and a choice of the initial configuration of the pump,
$H_-$. This dependence is suppressed from our notation because
Thm.~\ref{BPT} implies that the dependence disappears in the
limit. Ultimately, this independence is a consequence of propagation
estimates.

Eq.~(\ref{MOURRE}) implies a minimal escape
velocity estimate \cite{DG,HSS} for the autonomous dynamics
generated by $H(s)$. The constants involved are understood as being
uniform in $s \in I$ and in the stated range for $a$. Within proofs
we shall abbreviate $H\equiv H(s)$.

\begin{lem}\label{FIRSTL_PRE}
Let $\chi$ in $C_0^\infty(0,\infty)$ (in particular with support
away from $E=0$). Then, for some $\theta >0$, for all $a \in \R$,
$b,t \ge 0$ and $N \in \N$:
\begin{equation}\label{}
\|F(A\le a-b+ \theta t) \e^{-\i H(s) t}\chi(H(s))F(A\ge a)\| \le
C_N(\theta)(b+\theta t)^{-N}\;.
\end{equation}
Similarly, if $b,t <0$, then
\begin{equation}
\|F(A\ge a-b+\theta t)\e^{-\i H(s) t}\chi(H(s))F(A\le a)\| \le
C_N(\theta) |b+\theta t|^{-N}\;. \label{eq:prop-}
\end{equation}
\end{lem}

\begin{proof}[{\bf Proof.}]
The case of $b=0$ is covered by Thm.~1.1 in \cite{HSS}, since its
hypothesis (besides of (\ref{MOURRE})) that $\ad{k}{A}{f(H)}$ is bounded
for $f\in C_0^\infty(\R)$ and $k\ge 1$ holds true by (\ref{eq:mc}).
To be precise, the result is formulated there for
$\supp\chi\subset\Delta$, where $\Delta$ is as in (\ref{MOURRE}) and
$0 < \theta < \theta_0$, but it extends to our case by a covering
argument. Actually, the proof
given there essentially covers the general case $b\neq 0$. More
precisely, let
\begin{equation}
A_{t\tau}= \tau^{-1}(A-a + \frac{b}{2} -\theta_0t) \nonumber
\end{equation}
and $f \in C^\infty(\R)$ be a function with
$|f^{(k)}(x)| \le C \langle x\rangle^{-k}, f' \le 0$, and $f(x) = 0$ for
$x \ge 0$. Then
\begin{equation}
\|f(A_{t\tau})\e^{- \i Ht} \chi(H)F(A\ge a)\| \le C_N \tau^{-N} \nonumber
\end{equation}
uniformly in $0 \le t\le \tau$ and $a \in \R$. For $b = 0$, this is
equation (2.11) in \cite{HSS}, whose proof applies to $b \ge 0$ as
well. Let $\tau=b+t$. Since
\begin{equation}
\frac{b-\theta t}{b+t} \ge \frac{b/2 - \theta_0 t}{b+t} + \delta \nonumber
\end{equation}
for some $\delta >0$ and all $b,\,t \ge 0$ we have
\begin{equation}
F(A \le a -b + \theta t) \le
F\Bigl(\frac{A-a +(b/2)-\theta_0 t}{b+t} \le-\delta\Bigr)
\le f(A_{t\tau}) \nonumber
\end{equation}
for some $f$ of the required type.
\end{proof}

The lemma after next is a refined version of Prop.~\ref{Itraceclass} 
above. The discussion is 
simplified by the observation that the regularized currents 
Eq.~(\ref{eq:cu1}) may be written as
\begin{equation}
I_{j\pm}(s,a)=\chi(H(s))\i[H_b(s), f(\pm A_j-a)]\chi(H(s))\;,
\label{eq:cu2}
\end{equation}
where $H_b(s)=H(s)b(H(s))$ with $b\in C_0^\infty(\R)$ and
$b\chi=\chi$. The unregularized current, Eq.~(\ref{eq:cu}), is independent 
of $s$ by A3
and the commutation $[H(s), \Pi_j]=0$.
Since the commutation fails when $H(s)$ is replaced by $H_b(s)$
the regularized current is a-priori epoch dependent. However, the
commutation is essentially recovered on the ``box'' $f(\pm A-a)$
when $a$ is large. More precisely:
\begin{lem}\label{SECONDL_PREbis} For $j=0,\ldots n$ and $a\ge -1$ we have
\begin{equation}
\|[H_b(s), \Pi_j]f(\pm A-a)(\pm A-a+\i)^2\|\le C
\label{eq:comm1}
\end{equation}
\end{lem}
\begin{proof}[{\bf Proof.}]
Since $(A+\i)^{-2}f(\pm A-a)(\pm A-a+\i)^2$ is uniformly bounded in
$a\ge 1$ we may prove instead that
\begin{equation}
\|[H_b(s), \Pi_j](A+\i)^2\|\le C\;.
\label{eq:comm2}
\end{equation}
We begin by showing that $\Pi_iH_b\Pi_j(A+\i)^2=\Pi_iH_b\Pi_j(A_j+\i)^2$
is bounded for
$i\neq j$. Indeed,
\begin{equation*}
H_b(A_j+\i)^2=(A_j+\i)^2H_b+2(A_j+\i)[H_b,A_j]+[[H_b,A_j],A_j]\;,
\end{equation*}
from which we infer that $(A_j+\i)^{-2}H_b(A_j+\i)^2$ is bounded. Hence
so is (use $\Pi_iA_j=0$)
\begin{equation*}
\Pi_iH_b\Pi_j(A_j+\i)^2=-\Pi_i(A_j+\i)^{-2}H_b(A_j+\i)^2\Pi_j\;.
\end{equation*}
Now, setting $\bar\Pi_j=1-\Pi_j=\sum_{k=0,k\neq j}^n\Pi_k$, we have
\begin{equation*}
[H_b, \Pi_j](A+\i)^2=(\bar\Pi_j H_b\Pi_j-\Pi_jH_b\bar\Pi_j)(A+\i)^2\;,
\end{equation*}
and the result follows.
\end{proof}
This implies that although the regularized current operator a-priori
depends on the epoch and the pump, the dependence disappears in
the limit of large $a$. 
\begin{lem}\label{SECONDL_PRE}
\begin{eqnarray}
\|\i[H_b(s), f(\pm A_j-a)]\chi(H(s))\|_1&\le& C\;,
\label{eq:cutr}\\
\|F(A<\pm a-\alpha)\i[H_b(s), f(\pm A_j-a)]\chi(H(s))\|_1&\le&
C_N(1+\alpha)^{-N}\;,
\label{eq:cu<}\\
\|F(A>\pm a+\alpha)\i[H_b(s), f(\pm A_j-a)]\chi(H(s))\|_1&\le&
C_N(1+\alpha)^{-N}
\label{eq:cu>}
\end{eqnarray}
for $a,\,\alpha\ge 1$. The same bounds hold in operator norm if
$\chi(H(s))$ is replaced with $(\pm A- a+\i)^2$.
\end{lem}
The estimate (\ref{eq:cu>})
prevents the current operators from being located very far in the
outgoing region of phase space. This will play a role in the next
lemma and in Sect.~\ref{sect:3}. Actually, instead of
(\ref{eq:cu>})${}_-$ we shall use there the weaker statement with
characteristic function $F(A> a+\alpha)$. The pair
(\ref{eq:cu<})${}_+$ and (\ref{eq:cu>})${}_-$ keeps the current
operators away from the pump, a property used in
Sect.~\ref{sect:4}.
\begin{proof}[{\bf Proof.}]
The estimates to be proven are of
the form
\begin{align}
\|T[H_b, f(\pm A_j-a)]\chi(H)\|_1&\le
\|T[H_b, f(\pm A_j-a)](\pm A- a+\i)^2\|\|(\pm A- a+\i)^{-2}\chi(H)\|_1
\nonumber\\
&\le C\|T[H_b, f(\pm A_j-a)](\pm A-a+\i)^2\|\;,
\label{eq:step1}
\end{align}
where we used (\ref{eq:a5}).

We then have to establish the corresponding bounds for the remaining
operator norm. Since $g(A_j)=g(0)+(g(A)-g(0))\Pi_j$ we have
$f(\pm A_j-a)=f(\pm A-a)\Pi_j$ due to $f(-a)=0$ for $a>1$. Writing
$f=f\tilde f$, where $\tilde f(\cdot)=f(\cdot+2)$, the commutator
in (\ref{eq:step1}) is
\begin{equation*}
[H_b, f\Pi_j\tilde f]=
[H_b, f]\Pi_j\tilde f
+f\Pi_j[H_b,\tilde f]
+f[H_b,\Pi_j]\tilde f\;,
\end{equation*}
with $\tilde f=\tilde f(\pm A-a)$. In the contribution
to (\ref{eq:step1}) of the first two terms
the projections $\Pi_j$ may be
moved out to the right or to the left, using $[T,\Pi_j]=0$. As for the
last term, we use (\ref{eq:comm1}). At this point,
(\ref{eq:step1}) reduces to corresponding estimates for
\begin{equation}
\|T[H_b, f(A)](A+\i)^2\|+ \|T f(A)\|\;,
\label{eq:step2}
\end{equation}
where we replaced $\pm A-a$ by $A$, as the argument given below
applies to the more general case.\\

In the case of (\ref{eq:cutr}), where $T=1$, the second term
(\ref{eq:step2}) is clearly bounded. On the first term we use the commutator
expansion, see \cite{HS}, eq.~(B.16), or \cite{D},
based on the Helffer-Sj\"ostrand representation of $f(A)$.
\begin{eqnarray}
[H_b,f(A)]&=&
\sum_{k=1}^{m-1}\frac{1}{k!} (-1)^{k-1}\ad{k}{A}{H_b}f^{(k)}(A)
+R_m\;,
\label{eq:cexp}\\
R_m&=&-\frac{1}{\pi}\int dx\,dy\partial_{\bar z}\tilde f(z)
(A-z)^{-1}\ad{m}{A}{H_b}(A-z)^{-m}\;,\nonumber
\end{eqnarray}
where $\partial_{\bar z}=(\partial_x+\i\partial_ y)/2$ and $\tilde f$ is
an almost analytic extension of $f$, which may be chosen in such a way
that
\begin{equation*}
\int dx\,dy|\partial_{\bar z}\tilde f(z)||y|^{-p-1}
\le C\sum_{k=0}^{m+2}\|f^{(k)}\|_{k-p-1}
\end{equation*}
for $p=0,\ldots n$, the norms being defined as $\|f\|_k=\int dx \langle
x\rangle^k|f(x)|$. The choice of \cite{HS} is such that if $\supp f'$ is
compact, as it is in our case, then $|y|\ge C_1|x|-C_2,\,(C_1,\,C_2>0)$,
for $z=x+\i y\in\supp\tilde f$, which thus implies
\begin{equation}
\|(A-z)^{-1}(A+\i)\|\le C(|y|^{-1}+1)
\label{eq:b0}
\end{equation}
The expanded terms remain bounded if multiplied by $(A+\i)^2$ from the
right. For the remainder we obtain
\begin{equation}
\|R_m(A+\i)^2\|\le C\|\ad{m}{A}{H_b}\|\sum_{k=0}^{m+2}
(\|f^{(k)}\|_{k-m-1}+\|f^{(k)}\|_{k-m+1})\;,
\label{eq:remest}
\end{equation}
which is finite for $m\ge 3$. \\

We shall next prove (\ref{eq:cu<}) with $F(A< -\alpha)$ replaced
by $F(A< -3\alpha)$ for easing notion below. This may be further
replaced by $f(-\alpha^{-1}A-2)$ because of
$F(A< -3\alpha)=F(A< -3\alpha)f(-\alpha^{-1}A-2)$. Since
$f(-\alpha^{-1}A-2)f(A)=0$, the claim Eq.~(\ref{eq:step2}) reads
\begin{equation}
\|[H_b, f(-\alpha^{-1}A-2)]f(A)(A+\i)^2\|\le C_N\alpha^{-N}\;.
\end{equation}
We now apply the expansion (\ref{eq:cexp}) to $-\alpha^{-1}A-2$ instead of
$A$. Since $f^{(k)}(-\alpha^{-1}A-2)f(A)=0$ only the remainder
contributes. To bound $\|R_m(A+\i)^2\|$ we use
$\|(-\alpha^{-1}A-2-z)^{-1}(A+\i)\|\le C\alpha(|y|^{-1}+1)$ instead of
(\ref{eq:b0}) and collect the powers of $\alpha^{-1}$ generated by
each commutator. We so obtain the bound of (\ref{eq:remest}) times
$\alpha^{-m+2}$.\\

Finally, (\ref{eq:cu>}) can be brought into a form similar to
(\ref{eq:cu<}) by replacing $f$ with $f-1$. Both functions have the same
commutator with $H_b$ but essentially complementary supports.
\end{proof}
In Sect.~\ref{sect:3} we shall describe the dynamics
(\ref{SCHRO_EQ}) in terms of the autonomous dynamics generated by
the Hamiltonians $H(s)$. Once this is achieved, the choice of an
argument in initial 1-particle density matrix $\rho$, be it $H_-$
or $H(s)$, does not matter much for the current measurement, as
defined by (\ref{eq:cu1}). This is the content of the following
lemma.
\begin{lem} Let $\rho,\,\chi$ be as in Theorem~\ref{BPT}. Then, for
$a\ge 1$,
\begin{equation}
\lim_{t\to -\infty}\|(\rho(H(s))-\rho(H_-))\e^{-\i H(s)t}I_\pm(s,a)\|_1
=0\;.
\label{eq:init}
\end{equation}
\end{lem}
We remark that no claim of uniformity w.r.t. $s,\,a$ is
made here. The statement for $t\to+\infty$ is also true, but not needed.
\begin{proof}[{\bf Proof.}]
We first consider the case where $\rho$ is smooth, and since
$H$ is bounded below we may assume $\rho\in C_0^\infty(\R)$. Then
\begin{equation}
\rho(H)-\rho(H_-)\in\JJ_1\;
\label{eq:trcl}
\end{equation}
by Eq.~(\ref{eq:a1}).
To estimate the trace norm in (\ref{eq:init}) we use (\ref{eq:cu2}),
insert
$1=F(A \le a + \alpha)+F(A > a + \alpha)$ to the left of the
commutator in (\ref{eq:cu1}), and use (\ref{eq:cutr}, \ref{eq:cu>}).
We so obtain the bound
\begin{equation}
\|(\rho(H)-\rho(H_-))\e^{-\i Ht}\chi(H)F(A \le a +
\alpha)\|\cdot C+C_N\alpha^{-N}\;.
\label{eq:init1}
\end{equation}
The remaining operator norm is bounded for $t\le 0$ by
\begin{equation}
\|(\rho(H)-\rho(H_-))F(A \le a + \alpha+\theta t)\|+ C_N t^{-N}\;,
\label{eq:init2}
\end{equation}
where we used (\ref{eq:prop-}) with $b=0$. We pick $\alpha=-\theta t/2$,
so that the remainder in (\ref{eq:init1}) tends to zero, and
$\slim_{t\to-\infty}F(\ldots)=0$ in (\ref{eq:init2}).
Since $\rho(H)-\rho(H_-)$ is compact, the norm vanishes in the
limit.\\

In the general case, where $\rho$ is of bounded variation, let
$\rho_n\in C_0^\infty(\R),\,\sup_n\|\rho_n\|_\infty<\infty$, be a sequence
such that $\rho_n(\lambda)\to\rho(\lambda)$ pointwise, whence
$\slim_{n\to\infty}\rho_n(H)=\rho(H)$, and the same for $H_-$ instead of
$H$.
Also,
$\slim_{t\to-\infty}(\e^{-\i Ht}P_{\rm ac}(H)-\e^{-\i H_-t}\Omega_-)=0$
by definition of the wave operator $\Omega_-=\Omega_-(H_-,H)$
\cite{RS3, Y}. Since $I_\pm (s,a)$ is trace class
by (\ref{eq:cutr}) and since
\begin{equation}
\slim_{n\to\infty}X_n=0\;,\quad Y\in\JJ_1\; \Rightarrow\;
\lim_{n\to\infty}\|X_nY\|_1=0\;,
\label{eq:impl}
\end{equation}
we have
\begin{eqnarray}
\lim_{n\to\infty}\|(\rho_n(H)-\rho(H))I_\pm(s,a)\|_1&=&0\;,
\nonumber\\
\lim_{n\to\infty}\|(\rho_n(H_-)-\rho(H_-))\Omega_-I_\pm(s,a)\|_1&=&0\;,
\nonumber\\
\lim_{t\to-\infty}
\|(\e^{-\i Ht}-\e^{-\i H_-t}\Omega_-)I_\pm(s,a)\|_1&=&0\;,
\label{eq:limwo}
\end{eqnarray}
where we dropped $P_{\rm ac}(H)$ due to $P_{\rm ac}(H)\chi(H)=\chi(H)$.
We then estimate ($I_\pm\equiv I_\pm(s,a)$)
\begin{eqnarray*}
\lefteqn{\|(\rho(H)-\rho(H_-))\e^{-\i Ht}I_\pm\|_1}\\
&\le&\|(\rho(H)-\rho_n(H))\e^{-\i Ht}I_\pm\|_1+
\|(\rho_n(H)-\rho_n(H_-))\e^{-\i Ht}I_\pm\|_1\\
&&+\|(\rho_n(H_-)-\rho(H_-))
(\e^{-\i Ht}-\e^{-\i H_-t}\Omega_-)I_\pm\|_1
+\|(\rho_n(H_-)-\rho(H_-))\e^{-\i H_-t}\Omega_-I_\pm\|_1\\
&\le&\|(\rho(H)-\rho_n(H))I_\pm\|_1+
\|(\rho_n(H)-\rho_n(H_-))\e^{-\i Ht}I_\pm\|_1\\
&&+C\|(\e^{-\i Ht}-\e^{-\i H_-t}\Omega_-)I_\pm\|_1
+\|(\rho_n(H_-)-\rho(H_-))\Omega_-I_\pm\|_1
\end{eqnarray*}
Given $\epsilon>0$ we first pick $n$ large enough such that the first
and the last term together are less than $\epsilon/2$. For large
negative $t$ the same is true for the second and the third term by
(\ref{eq:limwo}) and the first part of the proof.
\end{proof}
\section{The adiabatic limit}\label{sect:3}
The current generated by adiabatic pumps can be understood
within the general framework of the theory of linear response: The
adiabatic limit corresponds to weak driving, and the formal
derivation of Eq.~(\ref{eq:bpt}) in \cite{BPT} is a perturbation
expansion. Formally, the change in the state $\rho$ of system,
obtained by linearizing the Hamiltonian around epoch $s$ is
\begin{equation}
U_\varepsilon(s,s_-)\rho(H(s)) U_\varepsilon(s_-,s) - \rho(H(s))\approx
-\i\varepsilon\int_{-\infty}^0 dt\, t\, \e^{ \i H(s)t}[\dot{H}(s),\rho(H(s))]
\e^{- \i H(s)t}.\label{heuristic}
\end{equation}
One immediate difficulty with this expression is that the
integrand on the right hand side grows linearly with time. As an
operator identity the above equation does not make sense, and the
right hand side is not recognizably of order $\varepsilon$.

One of the reasons why a perturbation expansion can nevertheless
be made is that the current of Eq.~(\ref{eq:cu1}) has good
localization in phase-space and so distinguishes a region of phase
space where $\rho$ is to be evaluated. Therefore, only a
restricted range of times contribute to the integral: The time
associated with propagation from the pump to the ammeter. This
makes the expectation value of the current well defined. Estimates
of this kind are known as propagation estimates and are controlled
by Mourre theory.

In this section the limit $\varepsilon\to 0$ in (\ref{eq:bpt2}) is taken
as $a$ is kept fixed. The main result is the following.
\begin{prop}\label{MAIN_S4}
For fixed $a\ge 1$, we have
\begin{equation}
\lim_{\varepsilon \to 0} \varepsilon^{-1}
\tr \bigl( [U_\varepsilon(s,s_-)
\rho(H_-)U_\varepsilon(s_-,s)-\rho(H(s))]I_\pm(s,a)\bigr)=
\tr ( [\Omega_-^{(1)}(s),\rho(H(s))] I_\pm(s,a) )
\label{eq:adlim}
\end{equation}
uniformly in $s \in I$, where
\begin{equation}\label{WAVEOP_EXP1M}
\Omega_-^{(1)}(s) = - \i \int_{-\infty}^0 dt\, t\, \e^{ \i H(s)t}
\dot{H}(s)\e^{- \i H(s)t} \;.
\end{equation}
Moreover,
\begin{multline}
\lim_{\varepsilon \to 0} \varepsilon^{-1}
\tr \bigl( U_\varepsilon(s,s_-)
\rho(H_-)U_\varepsilon(s_-,s)(I_+(s,a)+I_-(s,a))\bigr)=\\
\tr ( [\Omega_-^{(1)}(s),\rho(H(s))] (I_+(s,a)+I_-(s,a)))\;.
\label{eq:adlim1}
\end{multline}
\end{prop}
\noindent{\bf Remarks.}
\begin{list}{$\bullet$}{\setlength{\leftmargin}{\parindent}}
\item
The integral (\ref{WAVEOP_EXP1M}) is a trace class norm
convergent integral once multiplied from the left or from the right by
$I_\pm(s,a)$, as in (\ref{eq:adlim}).
\item
Eq.~(\ref{eq:adlim}) separately describes, to leading order, the
incoming and outgoing currents in an adiabatically evolved
state as compared to the corresponding instantaneous state
$\rho\big(H(s)\big)$.
Eq.~(\ref{eq:adlim1}) is then an immediate consequence of
(\ref{eq:stat}).
\item
$\Omega_-^{(1)}(s)$ is formally the first order in $\varepsilon$ term in the
expansion of the wave operator which for fixed $s$ relates
$U_\varepsilon(s',s)$ to the autonomous dynamics $u_s(s'-s)$ generated
by the Hamiltonian $H(s)$:
\begin{equation*}
u_s(\sigma):=\e^{-iH(s)\varepsilon^{-1}\sigma}\;,
\end{equation*}
where $\varepsilon$ has been suppressed from the notation. In the next
section we shall also meet the first
order term in the expansion of the scattering operator (\ref{eq:dscatt0}) 
relating these two dynamics,
\begin{equation}\label{eq:dscatt}
S^{(1)}(s)=\Omega_-^{(1)}(s) -\Omega_+^{(1)}(s)
= -\i \int_{-\infty}^{\infty} dt\, t\, \e^{\i H(s)t}
\dot{H}(s)\e^{-\i H(s)t} \;.
\end{equation}
The reason
$\Omega_-^{(1)}(s)$, rather than $\Omega_+^{(1)}(s)$, appears in
(\ref{eq:adlim}) is that the initial condition was set in the past
of the current measurement.
\item
All estimates in this section hold true uniformly in
$s\in I,\,s_-\le 0$. Constants may however depend on $a$.
\end{list}
The two dynamics, $U_\varepsilon$ and $u_s$, are compared by means of
the Duhamel formula
\begin{equation}\label{DUHAMEL}
U_\varepsilon(s',s) = u_s(s'-s)+\i \varepsilon^{-1}
\int_{s'}^s ds_1 U_\varepsilon(s',s_1)
(H(s_1)-H(s))u_s(s_1-s)\;.
\end{equation}
Starting at epoch $s$, the Heisenberg dynamics of the currents
carries them through the scatterer within a finite time, i.e.,
essentially still at same epoch $s$. When applying (\ref{DUHAMEL})
to $I_\pm(s,a)$ it is thus appropriate to linearize $H(s_1)-H(s)$
around $s$. The next lemma essentially says that calculating the
current to first order in $\varepsilon$ the error one makes is
second order in $\varepsilon$:
\begin{lem}\label{FIRSTL_S4}
We have
\begin{equation}
\varepsilon^{-1}\| [U_\varepsilon(s_-,s) -
(u_s(s_--s)+X_\varepsilon(s_-,s))]I_\pm(s,a)\|_1 \le C \varepsilon\;,
\end{equation}
where
\begin{equation*}
X_\varepsilon(s_-,s)=\i \varepsilon^{-1}
\int_{s_-}^sds_1(s_1-s) U_\varepsilon(s_-,s_1)
\dot{H}(s) u_s(s_1-s)\;.
\end{equation*}
\end{lem}
In turn this yields the following expression for the (rescaled)
currents at epoch $s$:
\begin{eqnarray}\label{CURR_EXP}
\lefteqn{\varepsilon^{-1}\tr(\rho(H_-)U_\varepsilon(s_-,s)
I_\pm(s,a)U_\varepsilon(s,s_-))=}\hspace{4cm}\nonumber\\
&& \varepsilon^{-1}\tr(u_s(s-s_-)\rho(H_-)u_s(s_--s)I_\pm(s,a))+\nonumber\\
&& \varepsilon^{-1}
\tr(X_\varepsilon(s_-,s)^*\rho(H_-)u_s(s_--s)I_\pm(s,a))+\nonumber\\
&& \varepsilon^{-1}
\tr(u_s(s-s_-)\rho(H_-)X_\varepsilon(s_-,s)I_\pm(s,a))+\nonumber\\
&& \varepsilon^{-1}
\tr(\rho(H_-)X_\varepsilon(s_-,s)I_\pm(s,a)X_\varepsilon(s_-,s)^*)+
\O{\varepsilon}\;.
\end{eqnarray}
\begin{proof}[{\bf Proof.}]
Using Duhamel's formula (\ref{DUHAMEL}) and
\begin{eqnarray}
&& H(s_1)-H(s) -(s_1-s)\dot{H}(s) = ( H(s_1)-H(s)
-(s_1-s)\dot{H}(s))F(A=0)\;, \nonumber \\
&& \qquad \qquad \|H(s_1)-H(s)-(s_1-s)\dot{H}(s)\| \le C |s_1-s|^2\;,
\nonumber
\end{eqnarray}
see (\ref{eq:conf}), Assumption A1 and $\Pi_0=F(A=0)\Pi_0$, we
are left with showing that
\begin{equation}\label{PROP_EST1}
\varepsilon^{-2} \int_{s_-}^s ds_1\, |s_1-s|^2
\cdot \|F(A=0)u_s(s_1-s)I_\pm(s,a)\|_1 \le C\varepsilon\;.
\end{equation}
We insert $1=F(A \le a + \alpha)+F(A > a + \alpha)$ to the left of the
commutator in (\ref{eq:cu1}).
By (\ref{eq:cu2}, \ref{eq:cu>}) the trace norm appearing under the integral
is bounded as
\begin{multline*}
\|F(A=0)u_s(s_1-s)I_\pm(s,a)\|_1\\
\le C \|F(A=0)u_s(s_1-s)\chi(H(s))F(A \le a + \alpha)\| + C_N (1 +
\alpha)^{-N}\;,
\end{multline*}
where we also used (\ref{eq:cutr}). By (\ref{eq:prop-}) with $b=0$
the latter norm is estimated as
\begin{multline}
\| F(A=0) u_s(s_1-s)\chi(H(s))F(A\le a+\alpha)\|
\label{eq:dis}\\
\le C \|F(A=0) F(A\le a +\alpha + \theta \varepsilon^{-1}(s_1-s))\| +
C_N(1 + \varepsilon^{-1}|s_1-s|)^{-N}\;.
\end{multline}
Picking $\alpha = \frac{1}{2}\theta\varepsilon^{-1}(s-s_1) >0$, the
first term vanishes for $s-s_1 > 2a\theta^{-1} \varepsilon$. Now
Eq.~(\ref{PROP_EST1}) holds because the l.h.s. is estimated by
a constant times
\begin{equation}
\varepsilon^{-2} \int_0^\infty d\sigma\, \sigma^2 (F(\sigma\le
2a\theta^{-1}\varepsilon)+(1 + \varepsilon^{-1}\sigma)^{-N})
\le C(\varepsilon a^3+ \varepsilon)\;,
\label{eq:dis1}
\end{equation}
which proves the lemma.
\end{proof}
For later purposes we retain the following estimate from the above proof:
\begin{equation}
\| F(A=0) \e^{-\i H(s)t}I_\pm(s,a)\|
\le F(|t|\le 2a\theta^{-1})+C_N(1+|t|)^{-N}\;.
\label{eq:lp}
\end{equation}
In particular, it proves the first remark after Prop.~\ref{MAIN_S4}.\\

The last term in Eq.~(\ref{CURR_EXP}) is also $\O{\varepsilon}$,
as shown by the next estimate.
\begin{lem}
We have
\begin{equation}
\|X_\varepsilon(s_-,s)I_\pm(s,a)X_\varepsilon(s_-,s)^*\|_1
\le C\varepsilon^2\;.
\label{eq:rem}
\end{equation}
\end{lem}

\begin{proof}[{\bf Proof.}]
As in the proof of the previous lemma, the norm in (\ref{eq:rem}) is
bounded by
\begin{equation}
\varepsilon^{-2} \int_{s_-}^s ds_1\,ds_2 |s_1-s||s_2-s|
\|F(A=0)u_s(s_1-s)I_\pm(s,a)u_s(s_2-s)^*F(A=0)\|_1\;.
\label{eq:rem1}
\end{equation}
Using again (\ref{eq:cu2}, \ref{eq:cu>}), this last norm by
itself is bounded by
\begin{equation*}
\Bigl(\|F(A=0)u_s(s_1-s) \chi(H(s))F(A\le a + \alpha)\| +
C_N(1+\alpha)^{-N}\Bigr) \cdot \Bigl(s_1\to s_2\Bigr)\;
\end{equation*}
where $(s_1\to s_2)$ is shorthand for the previous expression with
$s_1$ replaced by $s_2$. We pick $\alpha = \frac{1}{2}\theta \varepsilon^{-1}
\min(s-s_1,s-s_2)\ge 0$. Then the previous expression is estimated as
\begin{equation*}
\Bigl(F(\sigma_1\le2a\theta^{-1}\varepsilon)+
(1+\frac{1}{2}\theta \varepsilon^{-1}\min(\sigma_1,
\sigma_2))^{-N}\Bigr)
\cdot \Bigl(s_1\to s_2\Bigr)\;,
\end{equation*}
where $\sigma_i=s-s_i$. Using
\begin{eqnarray*}
\int_0^\infty d\sigma_2\,\sigma_2
(1+\frac{1}{2}\theta \varepsilon^{-1}\min(\sigma_1,\sigma_2))^{-N}
\le C(\sigma_1^2+\varepsilon^2)\;,\\
\int_0^\infty d\sigma_1d\sigma_2\,\sigma_1\sigma_2
(1+\frac{1}{2}\theta \varepsilon^{-1}\min(\sigma_1,\sigma_2))^{-N}
\le C\varepsilon^4\;,
\end{eqnarray*}
the claimed bound is established for (\ref{eq:rem1}).
\end{proof}
We now turn to the second and third term in Eq.~(\ref{CURR_EXP}), which will
account for the current in the adiabatic limit.
\begin{lem}
\begin{equation}
\lim_{\varepsilon\to 0}
\|\bigl( U_\varepsilon(s_-,s_1) -u_s(s_--s_1)\bigr)
u_s(s_1-s)\rho(H(s))\chi(H(s))(\pm A-a+\i)^{-2}\|_1=0 \;,
\label{eq:devol}
\end{equation}
uniformly also in $s_1\in[s_-, s]$.
\end{lem}
\begin{proof}[{\bf Proof.}]
We first consider the case where $\rho$ is smooth. Since
$T=\chi(H(s))(\pm A-a+\i)^{-2}\in\JJ_1$ by (\ref{eq:a5}) and
$\slim_{\alpha\to\infty}F(A> \alpha)=0$ we have
$\lim_{\alpha\to\infty}\|F(A> \alpha)T\|_1=0$. By approximation we may
thus prove
\begin{equation*}
\lim_{\varepsilon\to 0}
\|( U_\varepsilon(s_-,s_1) -u_s(s_--s_1))
u_s(s_1-s)\rho(H(s))\tilde\chi(H(s))F(A\le \alpha)\|=0\;,
\end{equation*}
where $\tilde\chi\in C_0^\infty(0,\infty)$ with $\tilde\chi\chi=\chi$.
To estimate this operator norm we use (\ref{DUHAMEL}) together with
\begin{equation}
\|H(s_2)-H(s)\| \le C |s_2-s|
\label{eq:deltah}
\end{equation}
to obtain the bound
\begin{equation*}
\varepsilon^{-1} \int_{s_-}^{s_1} ds_2 |s_2-s|
\|F(A=0) u_s(s_2-s)\rho(H(s))\tilde\chi(H(s)) F(A\le \alpha)\|
\le C\varepsilon(1+\alpha^2)\;,
\end{equation*}
where we used, see the argument in connection with (\ref{eq:dis}), that
the norm under the integral is bounded by
$F(s-s_2\le\varepsilon\theta^{-1}\alpha)+C_N
(1 + \varepsilon^{-1}|s_2-s|)^{-N}$. The general case, where
$\rho$ is of bounded variation, is also dealt with by approximating
$\rho(H)=\slim_{n\to\infty}\rho_n(H)$ with $\rho_n$ smooth. In fact we can
pick $n$ so that $\|(\rho_n(H)-\rho(H))T\|_1$ is arbitrarily small
(uniformly in $s$).
\end{proof}
\begin{proof}[{\bf Proof} {\rm of Prop.~\ref{MAIN_S4}.}]
As mentioned in the introduction,
$U_\varepsilon(s,s_-)\rho(H_-)U_\varepsilon(s_-,s)$ is independent of
$s_-\le 0$. We may thus evaluate the r.h.s. of (\ref{CURR_EXP}) in the
limit $s_-\to\infty$. By estimate (\ref{eq:init}) and its adjoint, this
amounts to replacing $\rho(H_-)$ by $\rho(H(s))$, i.e.,
\begin{eqnarray}\label{CURR_EXP1}
\lefteqn{\varepsilon^{-1}\tr(\rho(H_-)U_\varepsilon(s_-,s)
I_\pm(s,a)U_\varepsilon(s,s_-))=
\varepsilon^{-1}\tr(\rho(H(s))I_\pm(s,a))+}\hspace{2cm}\nonumber\\
&& \varepsilon^{-1}\Lim_{s_-\to-\infty}
\tr(X_\varepsilon(s_-,s)^*u_s(s_--s)\rho(H(s))I_\pm(s,a))+\nonumber\\
&& \varepsilon^{-1}\Lim_{s_-\to-\infty}
\tr(u_s(s-s_-)X_\varepsilon(s_-,s)I_\pm(s,a)\rho(H(s)))+
\O{\varepsilon}\;,
\end{eqnarray}
where we also used $[u_s(s-s_-),\rho(H(s))]=0$ and (\ref{eq:rem}). The
first term one the r.h.s. is just the equilibrium value of the current
subtracted on the l.h.s. of Eq.~(\ref{eq:adlim}). As we are not going to
show that the limits in (\ref{CURR_EXP1}) exist, they are just to be
understood as sets of limit points.
We next claim that
\begin{equation}
\varepsilon^{-1}
\Lim_{s_-\to\infty}\tr((X_\varepsilon(s_-,s)^*u_s(s_--s)
-\varepsilon\Omega_-^{(1)}(s))
\rho(H(s))I_\pm(s,a))\to 0\;,\qquad(\varepsilon \to 0)\;.
\label{eq:ltp}
\end{equation}
By taking the complex conjugate this implies
\begin{equation*}
\varepsilon^{-1}
\Lim_{s_-\to\infty}\tr(I_\pm(s,a)\rho(H(s))(u_s(s-s_-)X_\varepsilon(s_-,s)
+\varepsilon\Omega_-^{(1)}(s)))\to 0\;,\qquad(\varepsilon \to 0)\;,
\end{equation*}
where we used $\Omega_-^{(1)}(s)^*=-\Omega_-^{(1)}(s)$. Used together in
(\ref{CURR_EXP1}) they prove (\ref{eq:adlim}).\\

We next note that, by a change of variables,
\begin{equation*}
\varepsilon \Omega_-^{(1)} (s) =
- \i \varepsilon^{-1} \int_{- \infty}^s ds_1
(s_1-s) u_s(s-s_1) \dot{H}(s) u_s(s_1-s)\;.
\end{equation*}
As noted before, the integral is convergent in trace class
norm upon multiplication from either side with $I_\pm(s,a)$. For the
purpose of proving
(\ref{eq:ltp}) we may thus assume it to have $s_-$ as its lower limit of
integration. Then the trace there, including the factor
$\varepsilon^{-1}$ in front, equals
\begin{multline}
- \i \varepsilon^{-2}\int_{s_-}^s ds_1(s_1-s)\times\\ tr(
u_s(s-s_1) \dot{H}(s)\bigl(U_\varepsilon(s_1,s_-)u_s(s_--s)-u_s(s_1-s)\bigr)
\rho(H(s))I_\pm(s,a))\;.\nonumber
\end{multline}
We use
\begin{equation*}
U_\varepsilon(s_1,s_-)u_s(s_--s)-u_s(s_1-s)
=U_\varepsilon(s_1,s_-)\bigl(u_s(s_--s_1)-U_\varepsilon(s_-,s_1)\bigr)
u_s(s_1-s)
\end{equation*}
and turn $u_s(s-s_1) \dot{H}(s)$ around the trace, so that the previous
expression estimated as
\begin{multline}
\varepsilon^{-2}\!\int_{s_-}^s\! ds_1|s_1-s|
\|\bigl(U_\varepsilon(s_-,s_1)-u_s(s_--s_1)\bigr)
u_s(s_1-s)\rho(H(s))\chi(H(s))(\pm A-a+\i)^{-2}\|_1\times\\
\|(\pm A-a+\i)^2[H_b(s), f(\pm
A-a)]\chi(H(s))u_s(s-s_1)F(A=0)\|\;.\nonumber
\end{multline}
The first factor tends to zero uniformly as $\varepsilon\to 0$ by
(\ref{eq:devol}),
so we are left to show
\begin{equation}
\varepsilon^{-2}\int_{s_-}^s ds_1|s_1-s|
\|F(A=0)u_s(s_1-s)\chi(H(s))[H_b(s), f(\pm A-a)](\pm A-a+\i)^2\|\le C\;.
\label{eq:dis2}
\end{equation}
We insert
$1=F(A \le a + \alpha)+F(A >a + \alpha)$ to the left of the
commutator. By (\ref{eq:cutr}, \ref{eq:cu>}) and the remark following
them we obtain the bound
\begin{equation*}
\varepsilon^{-2}\int_{s_-}^s ds_1|s_1-s|\bigl(
\|F(A=0)u_s(s_1-s)\chi(H(s))F(A \le a + \alpha)\|
+C_N(1+\alpha)^{-N}\bigr)\;,
\end{equation*}
where we take $\alpha = \frac{1}{2}\theta\varepsilon^{-1}(s-s_1) >0$ as
in (\ref{eq:dis}). The resulting bound differs from (\ref{eq:dis1}) by
having $\sigma$ instead of $\sigma^2$. Hence the bound (\ref{eq:dis2}).
\end{proof}
\section{The scattering limit}\label{sect:4}
In the previous section we saw that the currents can be computed from
frozen data in the adiabatic limit. These data were not directly
related to the frozen scattering data and involved objects like $H(s)$
and $\Omega_-^{(1)}(s)$. In this section we show that in the limit that
the ammeter is far, $a\to\infty$, then all we need to know is the frozen,
scattering operator $S$ and the initial state $\rho$.

First we show that in the large $a$ limit
the incoming, resp. outgoing currents have no scattering in the past,
resp. the future. As usual, all statements are uniform in $s \in I$.
\begin{lem}\label{sc1}
We have
\begin{eqnarray*}
&& \lim_{a \to \infty} \| \Omega_-^{(1)}(s) I_-(s,a) \|_1 = 0\;, \\
&& \lim_{a \to \infty}\|(\Omega_-^{(1)}(s)-S^{(1)}(s))I_+(s,a)\|_1 = 0\;.
\end{eqnarray*}
\end{lem}
We recall that $S^{(1)}$ was defined in (\ref{eq:dscatt}). These facts
will yield a first expression for the scattering limit of the current,
Eq.~(\ref{eq:adlim1}).
\begin{lem}\label{sc2}
We have
\begin{equation}
\lim_{a \to \infty}\tr ( [\Omega_-^{(1)}(s),\rho(H(s))] (I_+(s,a)+I_-(s,a)))
=\tr \bigl([S^{(1)}(s),\rho(H(s))]I_+(s,a)\bigr)\;,
\label{eq:sclim3}
\end{equation}
where $a$ on the r.h.s. is arbitrary.
\end{lem}
We then express the latter result in terms
of ``frozen'' scattering data, such as the
scattering operator $S(s)$, defined in (\ref{eq:scatt}). Notice that it
acts on the
asymptotic Hilbert space $L^2(\R_+,\C^n)$ of the channels, rather than to
$\HH$. Further distinguished operators acting there are the Neumann
Hamiltonian $H_0$, introduced in the introduction, and the generator of
dilations,
\begin{equation*}
A_0 =\frac{1}{2\i}(\frac{d}{dx}x+x\frac{d}{dx})\;,
\end{equation*}
which may be regarded as a model for (\ref{eq:gdil}). The trace
(\ref{eq:sclim3}) can then finally be computed exactly.
\begin{prop}\label{sc3}
Suppose, in addition to the hypotheses of
Thm.~\ref{BPT}, that $\rho$ is smooth. Then,
\begin{eqnarray}
\tr \bigl([S^{(1)}(s),\rho(H(s))]I_+(s,a)\bigr)
&=&-\i \tr\bigl(\dot{S}(s)S(s)^*\rho'(H_0)\Pi_j\i[H_0, f(A_0-a)]\bigr)
\label{eq:sclim4}\\
&=&-\frac{\i}{2\pi}\int_0^\infty
d\rho(E)\Bigl(\frac{d S}{d s}S^*\Bigr)_{jj}\;,
\label{eq:sclim5}
\end{eqnarray}
where $A_{0j}=A_0\Pi_j$, and the scattering
operator $S(s)$ is defined in Eq.~(\ref{eq:scatt}).
\end{prop}
For $\rho$ which is not smooth an approximation argument will complete
the task.
\begin{proof}[{\bf Proof} {\rm of Lemma~\ref{sc1}.}]
Both claims of this proposition are the consequence of the stronger
bound:
\begin{equation*}
\| \Omega_\pm^{(1)}(s) I_\pm(s,a) \|_1 \le C_N a^{-(N-2)}\;,
\end{equation*}
for large enough $N$. Let us proof this bound for a case
of $\Omega_-$, the proof for $\Omega_+$ follows the same lines.
It is clear from Eq.~(\ref{WAVEOP_EXP1M}) that we may establish that
bound for
\begin{equation*}
\int^0_{-\infty}dt\, |t| \cdot
\|F(A=0)\e^{-\i H(s)t} I_-(s,a)\|_1\;.
\end{equation*}

The norm under this integral also appeared under the integral
(\ref{PROP_EST1}). We estimate it similarly by means of
(\ref{eq:cu2}, \ref{eq:cu>}), except that we insert $F(A\le -a +\alpha)$
(and the complementary projection) instead of $F(A\le a+\alpha)$.
By (\ref{eq:prop-}) we get
\begin{eqnarray*}
\|F(A=0)\e^{-\i H(s)t} I_-(s,a)\|_1 &\le &C \|F(A=0)F(A \le
-a+\alpha -b + \theta t)\| \\
&& + C_N(|b|+ |t|)^{-N} + C_N \alpha^{-N}\;.
\end{eqnarray*}
Choosing $b=-a/2,\, \alpha = (a/2 - \theta t)/2>0$, we
see that the
first term vanishes and we obtain the desired bound since
\begin{equation*}
\int_{-\infty}^0 dt\,|t| (\frac{a}{2}+\theta |t|)^{-N} \le C_N a^{-(N-2)}\;.
\end{equation*}
\end{proof}
An ingredient to the proof of Prop.~\ref{sc3} and, to a minor extent, of
Lemma~\ref{sc2} are the relations Eqs.~(\ref{eq:s1}, \ref{eq:s2}). 
Formally, Eq.~(\ref{eq:s1}) follows for
$\rho(\lambda) = \e^{\i\lambda t},\, (t \in \R)$, by a change of
variables in (\ref{eq:dscatt0}) and hence for general functions $\rho$.
For our purposes we shall need a somewhat more precise statement, given
by the first part of the following lemma (cf. the first remark after
Prop.~\ref{MAIN_S4}).
\begin{lem}\label{sc5}
(a) Eq.~(\ref{eq:s2}) and, for $\rho$ smooth, Eq.~(\ref{eq:s1}) hold true
if multiplied from either side by $I_+(s,a)$.\newline
(b) The wave operators $\Omega_\pm(s',s)=\slim_{T\to\pm\infty}
\e^{ \i H(s')T}\e^{- \i H(s)T}P_{\rm ac}(H(s))$ for the pair
$(H(s'), H(s))$ satisfy the intertwining property
\begin{equation}
f(H(s'))\Omega_\pm(s',s)=\Omega_\pm(s',s)f(H(s))
\label{eq:inter}
\end{equation}
for any (Borel) function $f$. \newline
(c) The scattering operators for the pairs $(H(s), H(s_i)),\,(i=1,2)$ are
related as
\begin{equation}
S(s,s_2)=\Omega_+(s_2,s_1)S(s,s_1)\Omega_-(s_2,s_1)^*\;.
\label{eq:scm}
\end{equation}

\end{lem}
\begin{proof}[{\bf Proof.}]
Parts (b, c) are standard \cite{RS3, AEGS3}. The wave operators exist
and are complete under our assumptions. The chain rule for wave operators,
\begin{equation*}
\Omega_\pm(s,s_1)=\Omega_\pm(s,s_2)\Omega_\pm(s_2,s_1)\;,
\end{equation*}
and $S(s,s_i)=\Omega_+(s,s_i)^*\Omega_-(s,s_i)$ imply
$S(s,s_1)=\Omega_+(s_2,s_1)^*S(s,s_2)\Omega_-(s_2,s_1)$. From this
(\ref{eq:scm}) follows by the completeness of scattering,
$\Omega_\pm(s_2,s_1)^*\Omega_\pm(s_2,s_1)=P_{\rm ac}(H(s_2))$.\\

We begin the proof of part (a) by claiming that
\begin{equation}\label{SECONDL_EXT}
\|(\Omega_+(s',s)-1)\chi(H(s))\| \le C |s'-s| \;,
\end{equation}
where $\chi\in C_0^\infty(0,\infty)$. Indeed, let
$\tilde\chi\in C_0^\infty(0,\infty)$ with $\tilde\chi\chi=\chi$.
Then, by (\ref{eq:inter}),
$\|(\Omega_+-1)\chi(H(s))\|\le
\|\tilde\chi(H(s'))(\Omega_+-1)\chi(H(s))\|+
\|\tilde\chi(H(s'))\chi(H(s))-\chi(H(s))\|$. The second term is bounded
by $\|\tilde\chi(H(s'))-\tilde\chi(H(s))\|$, and fits the bound
(\ref{SECONDL_EXT}) by (\ref{eq:a1}). For the first term we use the
fundamental theorem of calculus:
\begin{equation}
\tilde\chi(H(s'))(\Omega_+(s',s) -1)\chi(H(s)) =
\slim_{T \to \infty} \i \int_0^T dt\,
\tilde\chi(H(s'))\e^{ \i H(s')t}V \e^{- \i H(s)t}\chi(H(s))\;,
\label{eq:omega}
\end{equation}
where $V=H(s')-H(s)$. We write
$V=\langle A\rangle^{-r} V\langle A\rangle^{-r}$ and use (\ref{eq:deltah})
for $V$, as well the $H(s)$-smoothness of $\langle A\rangle^{-r}\chi(H(s))$
for $r>1/2$ (and similarly for $s'$), see item (ii) in
Sect.~\ref{sect:2}. An application of the Cauchy-Schwartz inequality on
matrix elements of (\ref{eq:omega}) yields (\ref{SECONDL_EXT}).\\

We can now prove the statement about Eq.~(\ref{eq:s2}), whose r.h.s. we
denote by $T$. Then $TI_+(s,a)$ is a convergent integral by
(\ref{eq:lp}). Moreover,
$T\e^{- \i H(s)t}I_+(s,a)=\e^{- \i H(s)t}TI_+(s,a)$, whence
$T\tilde\chi(H(s))I_+(s,a)=\tilde\chi(H(s))TI_+(s,a)$ for
$\tilde\chi\in C_0^\infty(0,\infty)$. Since
$\tilde\chi(H(s))I_+(s,a)=I_+(s,a)$ for $\tilde\chi\chi=\chi$ and $\chi$
as in (\ref{eq:cu1}), we may
thus prove
\begin{equation}
\tilde\chi(H(s))\partial_{s'}S(s',s)_{s'=s}I_+(s,a)=
-\i \tilde\chi(H(s))
\int_{-\infty}^\infty dt\, \e^{\i H(s)t}\dot{H}(s) \e^{-\i H(s)t}I_+(s,a)\;.
\label{eq:s3}
\end{equation}
We write
$S-1 = \Omega_+^*(\Omega_- - \Omega_+)$, so that
\begin{eqnarray*}
\tilde\chi(H(s))(S(s',s) -1)I_+(s,a) &=&
-\i \tilde\chi(H(s))\Omega_+^* \int_{-\infty}^\infty dt\,
\e^{ \i H(s')t} V \e^{- \i H(s)t}I_+(s,a) \\
&=& -\i \int_{-\infty}^\infty dt\,
\e^{\i H(s)t}\tilde\chi(H(s))(\Omega_+^* -1) V \e^{- \i H(s)t}
I_+(s,a)\\
&&- \i \tilde\chi(H(s))\int_{-\infty}^\infty dt\,
\e^{\i H(s)t}V \e^{- \i H(s)t}I_+(s,a)\;.
\end{eqnarray*}
The first integral is estimated as $C(a)|s'-s|^2$ due to
Eqs.~(\ref{SECONDL_EXT}, \ref{eq:deltah}, \ref{eq:lp}). In the second
one, the contribution coming from the remainder in
$V = \dot{H}(s)(s'-s) + O((s'-s)^2)$ is estimated the same way. This
implies (\ref{eq:s3}).\\

As for Eq.~(\ref{eq:s1}) in the case $\rho(\lambda) = \e^{\i\lambda t}$, the
change of variables mentioned before is legitimate, as the integrals are
convergent once multiplied with $I_+(s,a)$ due to (\ref{eq:lp}). The
result extends to $\rho(H(s))$ with $\rho\in C_0^\infty(\R)$ or
$\rho'\in C_0^\infty(\R)$, which amounts to the same since $H(s)$ is
bounded below.
\end{proof}
\begin{proof}[{\bf Proof} {\rm of Lemma~\ref{sc2}.}]
By Lemma~\ref{sc1} the l.h.s. of (\ref{eq:sclim3}) equals
$\lim_{a\to\infty}\tr \bigl([S^{(1)}(s),\rho(H(s))]$ $I_+(s,a)\bigr)$,
provided this limit exists. It does, since the expression is
independent of $a$.
For $\rho$ smooth this follows from Lemma~\ref{sc5}. Indeed, the
r.h.s. of (\ref{eq:s1}) commutes with $H(s)$, so that the independence
is seen as in (\ref{eq:indep}).
For general $\rho$ the conclusion
follows by approximation by a sequence $\{\rho_n\}$ of smooth
functions, such that $\slim\rho_n(H(s))=\rho(H(s))$. The traces converge
by (\ref{eq:impl}).
\end{proof}
\begin{proof}[{\bf Proof} {\rm of Prop.~\ref{sc3}.}]
For the sake of precision we recall that the scattering operator
$S(s)=\Omega_+(s)^*\Omega_-(s)$ depends on wave operators defined
in a two-Hilbert space setting:
$\Omega_\pm(s)=\slim_{T\to\pm\infty}\e^{\i H(s)T}J\e^{-\i H_0T}$,
where $J: L^2(\R_+,\C^n)\to \HH$ is the embedding given by
(\ref{eq:hs}). We may also replace $J$ by a smooth function $\tilde J$
on $\R$, which equals $0$ near $x=0$, and $1$ near $x=\infty$. This is
without effect on $\Omega_\pm(s)$, since
$\slim_{T\to\pm\infty}(J-\tilde J)\e^{-\i H_0T}=0$.
In particular, we may pick $\tilde J$ so that $\tilde J=1$ on $\supp v$,
see (\ref{eq:gdil}).
In the present context (\ref{eq:scm}) reads
$S(s',s)=\Omega_+(s)S(s')\Omega_-(s)^*$
and implies
\begin{equation*}
\partial_{s'}S(s',s)_{s'=s}=\Omega_+(s)\dot{S}(s)\Omega_-(s)^*\;,
\end{equation*}
where the differentiability is granted e.g. after multiplication with
$I_+(s,a)$.\\

We now prove Eq.~(\ref{eq:sclim4}) with $\Pi_j\i[H_0, f(A_0-a)]$
replaced by $\Pi_j\i[H_0, f(A-a)]=\i[H_0, f(A_j-a)]$.
Its l.h.s. equals, by Lemma~\ref{sc5},
\begin{eqnarray*}
\lefteqn{\tr\bigl(\Omega_+(s)\dot{S}(s)\Omega_-(s)^*\rho'(H(s))
\chi(H(s))I_+(a)\chi(H(s))\bigr)}\hspace{3cm}\\
&=&\tr\bigl(\dot{S}(s)S(s)^*\rho'(H_0)\chi(H_0)I_+(a)\chi(H_0)\bigr)\\
&&+\tr\bigl(\dot{S}(s)\Omega_-(s)^*\rho'(H(s))\chi(H(s))
[I_+(a), \Omega_+(s)]\chi(H_0)\bigr)\;.
\end{eqnarray*}
Here $I_+(a)$ is defined in Eq.~(\ref{eq:cu}). We turned
$\Omega_+(s)$ around the trace, which thereby moved from
$\HH$ to $L^2(\R_+,\C^n)$. The first term corresponds to the claim,
since $\chi\rho'\chi=\rho'$. It is again independent of $a$. The second
term will thus vanish as soon as it does in limit $a\to\infty$. The
commutator there is
$[I_+(a), \Omega_+(s)]=[I_+(a), \Omega_+(s)-\tilde J]$ by the above choice
$\tilde J$. We are so left to show that
\begin{equation*}
\lim_{a\to\infty}\|\chi(H(s))(\Omega_+(s)-\tilde J)\chi(H_0)I_+(a)\|_1=0\;,
\end{equation*}
and also with $\Omega_+(s)$ replaced by
$\Omega_+(s)^*$ and $\chi(H_0),\,\chi(H(s))$ interchanged. Using the
integral representation for $\Omega_+(s)-\tilde J$ the estimate reduces
to
\begin{multline*}
\int_0^\infty dt\,\|\chi(H(s)[H_0, \tilde J]\chi(H_0)
\e^{-\i H_0t}I_+(a)\|_1\\
\le C\int_0^\infty dt\,\|F(A=0)\chi(H_0)\e^{-\i H_0t}I_+(a)\|_1
\to 0\;,\qquad(a\to\infty)\;,
\end{multline*}
(and with $H_0,\,H(s)$ interchanged) which holds true by the estimates in
the proof of Lemma~\ref{sc1}.\\

We next replace $A$ by $A_0$ on the r.h.s. of
(\ref{eq:sclim4}). Since both traces are independent of $a$ we may, in
each of them, replace $f$ by a smeared switch function $\tilde f'$, such that
$\tilde f\in H^2(d)$ with $d>2$ (see the Appendix for notation). Since
$(\tilde f(A_j)-\tilde f(A_{0j}))\rho'(H_0)\in\JJ_1$ by
(\ref{eq:a4bis}), the difference is seen to vanish by cyclicity.\\

Finally, to prove (\ref{eq:sclim5}), we may keep
$f$ replaced by $\tilde f$ as before. It follows from
Eqs.~(\ref{eq:a4}, \ref{eq:app2}) that
\begin{multline*}
\tr\bigl(\dot{S}(s)S(s)^*\rho'(H_0) \Pi_j[H_0,
\tilde f(A_0-a)]\bigr) \\
= \tr\bigl(\dot{S}(s)S(s)^* \rho'(H_0) \Pi_j
(\tilde f(A_0-a-2\i)-\tilde f(A_0-a))H_0\bigr) \\
 = \frac{1}{2\pi}\int_0^\infty
\frac{dE}{2E}\rho'(E)E\Bigl(\frac{d S}{d s}S^*\Bigr)_{jj}
\cdot \int_{-\infty}^\infty d\lambda\
(\tilde f(\lambda-a-2\i)-\tilde f(\lambda-a))\;.
\end{multline*}
To compute the last integral, note that
$\int_{-\infty}^\infty d\lambda\, (\tilde f(\lambda-c)-\tilde
f(\lambda)) = -c$
for any $c \in \R$ which extends by analyticity to $|\Im c|<d$. Hence
the result.
\end{proof}
\begin{proof}[{\bf Proof} {\rm of Theorem \ref{BPT}.}]
We have shown that
\begin{equation}
\tr \bigl([S^{(1)}(s),\rho(H(s))]I_+(s,a)\bigr)
=-\frac{\i}{2\pi}\int_0^\infty
d\rho(E)\Bigl(\frac{d S}{d s}S^*\Bigr)_{jj}
\label{eq:pBPT}
\end{equation}
holds true if $\rho$ is smooth. If $\rho$ is of bounded variation
it can be approximated by a sequence $\rho_n\in C_0^\infty(\R)$
with $\rho_n(\lambda)\to\rho(\lambda)$ pointwise and uniformly
bounded total variation $\sup_nV(\rho_n)<\infty$. Then
$\slim_n\rho_n(H(s))=\rho(H(s))$
and $d\rho_n(\lambda)\to d\rho(\lambda)$ in the sense of
weak* convergence. Hence (\ref{eq:pBPT}) is inherited by the limit.
The proof is completed by Eqs.~(\ref{eq:adlim1}) and (\ref{eq:sclim3}).
\end{proof}
\appendix
\section{Appendix. Trace class properties of $g(H)f(A)$}\label{APP}
In this section we will discuss different properties of the operator
product $g(H(s))f(A)$. We first prove that operators of the
form $g(H_0)f(A_0)$
are Hilbert-Schmidt under suitable conditions on the functions $g$ and
$f$. For operator of this type one can
essentially write down the integral kernel, and use it in order to
compute the Hilbert-Schmidt norm. Heuristically the trace of an
operator is the integral over phase space of the
symbol (divided by $2\pi$). Since
\begin{equation*}
\frac{1}{2\pi}
\int dxdp=\frac{1}{2\pi}\int\frac{dE}{2E}da\;,
\end{equation*}
under the map $(x, p)\mapsto(E=p^2,a=px)$, we introduce the norms
\begin{equation*}
\|f\|^2=\int_{-\infty}^\infty da\,|f(a)|^2\;,\qquad
\dn g\dn^2=\int_0^\infty\frac{ dE}{2E}|g(E)|^2\;,
\end{equation*}
on functions $f$ on the real line, resp. $g$ on the half--line. The
Fourier transform is
$\hat f(t)=(2\pi)^{-1}\int\e^{-\i \lambda t}f(\lambda)d\lambda$.
\begin{prop}\label{FIRSTL_APP} We have
\begin{equation}
\|g(H_0)f(A_0)\|_2=(2\pi)^{-1/2}\dn g\dn\cdot\|f\|\;.
\label{eq:app1}
\end{equation}
If $g$ and $\hat f$ are continuous and the operator $g(H_0)f(A_0)$
is trace class, then
\begin{equation}
\tr\left(g(H_0)f(A_0)\right) =
\frac{1}{2\pi} \int_0^\infty \frac{dE}{2E}g(E) \cdot
\int_{-\infty}^\infty da\, f(a)\;.
\label{eq:app2}
\end{equation}
\end{prop}

\smallskip\noindent
\begin{proof}[{\bf Proof.}]
The space $L^2(\R_+)$ can be identified with the even
functions in $L^2(\R)$ by means of the map
$(J\varphi)(x)=\varphi(|x|)$,
with $\|J\varphi\|^2=2\|\varphi\|^2$. Consider the operator $g(p^2)f(A_0)$ on
$L^2(\mathbb R)$. Since it preserves parity and $p^2$ reduces to $H_0$ on
the (so identified) subspace of even functions (with projection $P_+$) we have
\begin{eqnarray}
\|g(H_0)f(A_0)\|_2^2&=&
\tr_{L^2(\R_+)}(\bar g(H_0)|f|^2(A_0)g(H_0))
\nonumber\\
&=&\frac{1}{2}\tr_{L^2(\R)}(P_+\bar g(p^2)|f|^2(A_0)g(p^2)P_+)
\nonumber\\
&=&\frac{1}{2}\tr_{L^2(\R)}(U^*P_+\bar g(p^2)|f|^2(A_0)g(p^2)P_+U)
\nonumber\\
&=&\frac{1}{2}\tr_{L^2(\R)}(P_+\bar g(x^2)|f|^2(-A_0)g(x^2)P_+)
\nonumber\\
&=&\tr_{L^2(\R_+)}(\bar g(x^2)|f|^2(-A_0)g(x^2))\;,\label{eq:app4}
\end{eqnarray}
where
$(U\psi)(k)=(2\pi)^{1/2}\hat{\psi}(k)$
is the Fourier
transform: $[U,P_+]=0,\,P_+=(1+U^2)/2$, $pU=-Ux$, and $A_0U=-A_0U$. Using
the Mellin transform $M:L^2(\R_+)\to L^2(\R)$,
\begin{equation*}
(M\varphi)(a)=(2\pi)^{-1/2}\int_0^\infty\frac{dx}{x^{1/2}} x^{-\i a}
\varphi(x)\;,
\end{equation*}
which diagonalizes $A_0=M^*aM$, one obtains that $h(a)$ has integral kernel
\begin{equation}
h(A)(x,y) = {(2\pi)}^{-1} (xy)^{-\frac{1}{2}} \hat{h}(\log \frac{y}{x})
\label{eq:intk}
\end{equation}
for $h \in L^1(\R)$. Since the kernel is continuous in $x,y >0$,
we can write for (\ref{eq:app4}) (cf. Thm.~3.9. \cite{SI})
\begin{equation*}
(2\pi)^{-1}\int_0^\infty\frac{dx}{x}|g(x^2)|^2
\int_{-\infty}^\infty da\,|f|^2(-a)\;.
\end{equation*}
By means of the change of variable $x^2=E$, this is the square of the
r.h.s. of (\ref{eq:app1}).\\

Using the Fourier transform as in (\ref{eq:app4}) we find
$\tr\bigl(g(H_0)f(A_0)\bigr) = \tr\bigl(g(x^2)f(-A_0)\bigr)$. The
integral kernel of the latter operator is obtained from (\ref{eq:intk})
and the trace is the integral over its diagonal, due to
Thm.~3.9 \cite{SI}.
\end{proof}
In the rest of this appendix we shall give an example of trace class
operator involving $A_0,\, H_0$ and use it to establish
$g(H(s))f(A)\in\JJ_1$ for a large enough class of operators.
\begin{lem} Let $g\in C_0^\infty(\R)$, then
\begin{eqnarray}
\|g(H(s))-g(H(s'))\|_1&\le& C|s-s'|\;,
\label{eq:a1}\\
g(H(s))-Jg(H_0)J^* &\in& \JJ_1 \;,
\label{eq:a2}
\end{eqnarray}
where $J: L^2(\R_+,\C^n)\to \HH$ is the embedding given by (\ref{eq:hs}).
\end{lem}
\begin{proof}[{\bf Proof.}]
We first prove (\ref{eq:a1}) for $g(\lambda)=(\lambda+\i)^{-2m}$, where
$m$ is as in Assumption A2. Setting $R(s)=(H(s)+\i)^{-1}$ we have by
(\ref{eq:conf}) and the resolvent identity
\begin{eqnarray*}
R(s)^{2m}-R(s')^{2m}&=&\sum_{k=1}^{2m}R(s)^{2m-k}(R(s)-R(s'))R(s')^{k-1}\\
&=&\sum_{k=1}^{2m}R(s)^{2m-k+1}\Pi_0(H(s')-H(s))\Pi_0R(s')^k\;.
\end{eqnarray*}
The desired bound now follows from A1 by using A2 either for $s$ or
$s'$. For general $g\in C_0^\infty(\R)$, as well as for the rest of this
proof, we use the Helffer-Sj\"ostrand representation,
\begin{equation}
g(H)=\frac{1}{\pi}\int_{\R^2}(H-z)^{-1}\partial_{\bar z}\tilde
g(z)dxdy\;,
\label{eq:a3}
\end{equation}
where $\partial_{\bar z}=(\partial_x+\i\partial_ y)/2$, and $\tilde g$ is
an almost analytic extension of $g$ with $\tilde g(z)$ vanishing to a
high power near the real axis. Before doing that, we set
$G(\lambda)=g(\lambda)(\lambda+\i)^{2m}$, so that the first term in
\begin{equation*}
g(H(s))-g(H(s'))=(R(s)^{2m}-R(s')^{2m})G(H(s'))
+R(s)^{2m}(G(H(s))-G(H(s'))
\end{equation*}
is taken care of. For the second, we apply (\ref{eq:a3}) to $G$ and are
led to estimate
\begin{multline*}
\|R(s)^{2m}[(H(s)-z)^{-1}-(H(s')-z)^{-1}]\|_1\le\\
\|(H(s)-z)^{-1}\|\|R(s)^{2m}\Pi_0\|_1\|H(s')-H(s)\|\|(H(s')-z)^{-1}\|\le
C|\Im z|^{-2}|s-s'|\;,
\end{multline*}
which completes the proof of (\ref{eq:a1}).\\

 Before proving (\ref{eq:a2}) we claim that
\begin{equation}
g(H)h\in\JJ_1\;,
\label{eq:a31}
\end{equation}
where $h$ acts as multiplication by $h\in C_0^\infty[0,\infty)$ on
$L^2(\R_+,\C^n)$ and by $h(0)$ on $\HH_0$. Since $g(H)\Pi_0\in\JJ_1$ it
will be enough to show $(1-\Pi_0)g(H)(1-\Pi_0)h\in\JJ_1$ or actually
just
$\|J(H-z)^{-1}J^*h(x)\|_1\le C|\Im z|^{-1}$. Since the kernel of
$(H-z)^{-1}$ is decaying exponentially \cite{CT}, matters are further reduced
to
\begin{equation*}
\|J_L(H-z)^{-1}J_L^*\|_1\le C|\Im z|^{-1}\;,
\end{equation*}
as an operator on $L^2([0,L],\C^n)$, where $J_L$ is the corresponding
embedding operator into $\HH$. The initial piece $[0,L]$ has to be taken
large enough, so that $\supp h\subset [0,L)$. Let now $B$ be the
quadratic form of the Bilaplacian,
\begin{equation}
B(\varphi,\psi)=\int_0^L\bar\varphi''(x)\psi''(x) dx\;,
\label{eq:bil}
\end{equation}
with domain given by the Sobolev space $W^2(0,L)$. We maintain that
\begin{gather*}
\|(1+B)^{1/2}J_L(H-z)^{-1}J_L^*\|\le C|\Im z|^{-1}\;,\\
(1+B)^{-1/2}\in \JJ_1\;,
\end{gather*}
which imply the claim. The first statement follows by A3 and
(\ref{eq:bil}) from
\begin{equation*}
J_L(H-\bar z)^{-1}J_L^*BJ_L(H-z)^{-1}J_L^*=T^*T\;,
\end{equation*}
with $T=1+zJ_L(H-z)^{-1}J_L^*$. The second statement can be seen by
computing the eigenvalues $k^4$ of the operator $B$ associated to
(\ref{eq:bil}). The latter is given as $B=d^4/dx^4$ with boundary
conditions $\varphi''=\varphi'''=0$ at $x=0,\,L$. From this one computes the
eigenvalues as the zeros of $1-\cosh kL\cos kL$.\\

To prove (\ref{eq:a2}) we make use of a smooth
embedding $\tilde J$ as in the proof of Prop.~\ref{sc3}. Since
$(J-\tilde J)g(H_0)\in \JJ_1$ (see \cite{RS2}, Thm. XI.20) we will
prove the trace class property for
\begin{equation*}
g(H)-\tilde Jg(H_0)\tilde J^*=g(H)(1-\tilde J\tilde J^*)
+(g(H)\tilde J-\tilde Jg(H_0))\tilde J^*\;.
\end{equation*}
The first term is trace class by (\ref{eq:a31}).
For the second term, Eq.~(\ref{eq:a3})
and the resolvent identity reduce matters to
\begin{equation*}
\|(H-z)^{-1}[\tilde J,H](H_0-z)^{-1}\tilde J^*\|_1\le
\|(H-z)^{-1}[\tilde J,H]\|\|h(x)(H_0-z)^{-1}\|_1\le
C|\Im z|^{-2}\;,
\end{equation*}
where $h\in C_0^\infty(\R)$ with $h\tilde J'=\tilde J'$.
\end{proof}
The Hardy class $H^2(d)$ consists of all functions $f$, analytic in the
strip $\{z\mid\vert\Im z\vert< d\}$, which satisfy
\begin{equation*}
\sup_{-d<y<d}\int_{-\infty}^\infty \vert f(x+\i y)
\vert^2dx<\infty\;.
\end{equation*}
We recall that
$f\in H^2(d)$ is also characterized by
$\sup_{0<y<d}\|\e^{y\vert k\vert}\hat f(k)\|_2<\infty$.
\begin{prop}[The pull-through formula]
For any $f\in H^2(d)$ with $d>2$
\begin{equation}
H_0 f(A_0) = f(A_0-2\i) H_0\;.
\label{eq:a4}
\end{equation}
\end{prop}
\begin{proof}[{\bf Proof.}] We first prove the statement for
$f(x)=\e^{-\i tx}$.
Under conjugation with the dilation operator,
$(\e^{-\i tA_0}\psi)(x)=\e^{-t/2}\psi(e^{-t}x)$,
the Neumann Laplacian becomes
\begin{equation}
\e^{\i tA_0}H_0\e^{-\i tA_0}=e^{-2t}H_0\;,
\label{eq:conj}
\end{equation}
hence $H_0\e^{-\i tA_0}=\e^{-\i t(A_0-2\i)}H_0$. Now, for $f\in H^2(d)$,
\begin{equation*}
H_0 f(A_0) = \int_{-\infty}^\infty H_0 \e^{\i tA_0} \hat f(t)dt
= \int_{-\infty}^\infty\e^{\i t(A_0-2\i)}H_0\hat f(t)dt
=f(A_0-2\i)H_0\;,
\end{equation*}
where the last step is justified by the above mentioned
property of the Hardy class functions.
\end{proof}
\begin{lem}\label{THIRDL_APP}
Let $f' \in H^2(d)$ with $d > 2$ and $g\in C_0^\infty(\R)$. Then
\begin{equation}
(f(A)-f(A_0)) g(H_0) \in J_1\;.
\label{eq:a4bis}
\end{equation}
\end{lem}
\begin{proof}
As mentioned, the Fourier transform of
$f'(\lambda)=\int_{-\infty}^\infty dt\, \e^{i\lambda t}\widehat{f'}(t)$
is bounded as
\begin{equation}\label{THIRDL_APP_EQ}
|\widehat{f'}(t)|\le C \e^{-y|t|}\;,
\end{equation}
for any $y<d$. We represent $f$ as
\begin{equation*}
f(\lambda) = f(0) + \int_{-\infty}^\infty dt\,
\Bigl(\frac{\e^{\i\lambda t}-1}{\i t}\Bigr) \widehat{f'}(t)\;,
\end{equation*}
and write
\begin{gather*}
\frac{1}{ \i t} (\e^{ \i At}-\e^{ \i A_0t}) =
\frac{1}{t} \int_0^t ds\, \e^{ \i A(t-s)}(A-A_0)\e^{ \i A_0s} \;,
\e^{\i A_0s}g(H_0) = g(H_0\e^{-2s})\e^{ \i A_0s} \;,
\end{gather*}
which follows from (\ref{eq:conj}). Applying Thm. XI.20~\cite{RS2} to
\begin{equation*}
(A-A_0)g(H_0 \e^{-2s}) = (w(x)p-\frac{i}{2}w'(x))g(H_0 \e^{-2s}) \;,
\end{equation*}
where $w(x)=v(x)-x$, one finds
\begin{equation}
\|(A-A_0)g(H_0\e^{-2s})\|_1 \le
C_\delta (\e^{s(\frac{3}{2}+\delta)}+\e^{s(\frac{1}{2}+\delta)})\;,
\end{equation}
for any $\delta > \frac{1}{2}$. We obtain
$\| t^{-1}(\e^{\i At}-\e^{\i A_0 t})g(H_0)\|_1 \le
C(1+\e^{\tilde{d} t})$
with $\tilde{d}= \frac{3}{2}+\delta$. Picking $\delta$ so that
$\tilde{d}<d$, we obtain the claim from Eq.~(\ref{THIRDL_APP_EQ}).
\end{proof}
The following example will be useful.
\begin{lem}
\begin{equation}
\Vert H_0(H_0+1)^{-2}(A_0-z)^{-2}\Vert_1\le C|\Im z|^{-2}\;.
\label{unper}
\end{equation}
\end{lem}
\begin{proof}[{\bf Proof.}]
It suffices to prove the claim for $|\Im z|$ large enough
since $\|(A_0-z+\i y)(A_0-z)^{-1}\|\le C|\Im z|^{-2}$ for small
$|\Im z|$. We shall do that for $|\Im z|>4$. By the
pull-through formula, Eq.~(\ref{eq:a4}), we compute
\begin{eqnarray*}
\lefteqn{H_0(A_0-z)^{-1}=
(A_0-z-2\i)^{-1}H_0}\hspace{0cm}\\
&&=(A_0-z-2\i)^{-1}(H_0^2+2H_0+1)H_0(H_0+1)^{-2}\\
&&=\bigl[H_0^2(A_0-z+2\i)^{-1} +
2H_0(A_0-z)^{-1} + (A_0-z-2\i)^{-1}\bigr]H_0(H_0+1)^{-2}\;.
\end{eqnarray*}
We multiply this expression from the left by $(H_0+1)^{-2}$ and from the
right by $(A_0-z)^{-2}$, so as to obtain the bound
\begin{equation*}
\Vert H_0(H_0+1)^{-2}(A_0-z)^{-2}\Vert_1\le
\sum_{k=-1}^1\Vert (A_0-z+2k\i)^{-1}H_0 (H_0+1)^{-2}(A_0-z_2)^{-1}
\Vert_1\;.
\end{equation*}
Eq.~(\ref{eq:app1}) now yields
$\|H_0^{1/2}(H_0+1)^{-1}(A_0-z)^{-1}\|_2=(2|\Im z|)^{-1/2}$. Hence
(\ref{unper}) follows from the H\"older inequality.
\end{proof}
\begin{lem} For $g\in C_0^\infty(0,\infty)$ we have
\begin{equation}
\sup_{a\in\R}\|(A-a\pm \i)^{-2}g(H)\|_1<\infty\;.
\label{eq:a5}
\end{equation}
In particular, for any $f\in C^\infty(\R)$ with $f(x)(x+\i)^2$ bounded,
\begin{equation}
f(A)g(H)\in\JJ_1\;.
\label{eq:a6}
\end{equation}
\end{lem}
\begin{proof}[{\bf Proof.}] Since $g(H_0)H_0^{-1}(H_0+1)^2$ is bounded,
by (\ref{unper})
the bound holds for $A_0,\,H_0$ in place of $A,\,H$. We first undo
the replacement in $A$ and write
\begin{equation*}
(A-a\pm \i)^{-2}g(H_0) = (A-a\pm \i)^{-2}\phi(x)g(H_0)+
(A-a\pm \i)^{-2}(1-\phi(x))g(H_0)\;,
\end{equation*}
where $\phi\in C_0^\infty[0,\infty)$ has $v(x)=x$ on $\supp(1-\phi)$.
Then $\phi(x)g(H_0) \in J_1$ (see \cite{RS2}, Thm. XI.20). For the
second term on the r.h.s. we write, dropping $a$,
$(A+\i)^{-2}(1-\phi(x))(A_0+\i)^2\cdot (A_0+\i)^{-2}g(H_0)$. Since the
last factor is trace class, we are left with showing that the first
factor is bounded. This follows from
\begin{equation*}
(1-\phi)(A_0+\i)^2=(A+\i)^2(1-\phi)-(A+\i)[\phi,A_0]-[\phi,A_0](A+\i)\;,
\end{equation*}
where $[\phi,A]= \i x\phi'(x)$ bounded. \\

The bound (\ref{eq:a5}) clearly also holds for
$(A+\i)^{-2}Jg(H_0)J^*=J(A+\i)^{-2}g(H_0)J^*$ and hence, by
(\ref{eq:a2}), as claimed.
\end{proof}
\noindent
{\bf Acknowledgments.} This work is supported in part by the ISF; the
Fund for promotion of research at the Technion; the Texas Advanced
Research Program and NSF Grant PHY-9971149, and the Swiss National
Science Foundation.
G.M.G. thanks T. Spencer for hospitality at the Institute for Advanced
Study, where part of this work was done.

\bibliographystyle{amsplain}

\end{document}